\documentclass[fleqn,usenatbib]{mnras}

\usepackage{newtxtext,newtxmath}
\usepackage{mwe,tikz}\usepackage[percent]{overpic}

\usepackage[T1]{fontenc}

\DeclareRobustCommand{\VAN}[3]{#2}
\let\VANthebibliography\thebibliography
\def\thebibliography{\DeclareRobustCommand{\VAN}[3]{##3}\VANthebibliography}

\usepackage{graphicx}	
\usepackage{amsmath}	
\usepackage{float}

\title[BPASS WDB and BHB]{Predicting gravitational wave signals from BPASS White Dwarf Binary and Black Hole Binary populations of a Milky Way-like galaxy model for LISA}

\author[P. Tang et al.]{
P. Tang,$^{1}$\thanks{E-mail: petra.tang@auckland.ac.nz}
J.J. Eldridge,$^{1}$
R. Meyer$^{2}$
A. Lamberts$^{3,4}$
G. Boileau$^{5}$
and W. G. J. van Zeist$^{6,7,1}$
\\
$^{1}$Department of Physics, University of Auckland, Private Bag 92019, Auckland, New Zealand\\
$^{2}$Department of Statistics, University of Auckland, Private Bag 92019, Auckland, New Zealand\\
$^{3}$Université Côte d’Azur, Observatoire de la Côte d’Azur, CNRS, Laboratoire Lagrange, Bd de l’Observatoire, F-06304 Nice, France\\
$^{4}$Université Côte d’Azur, Observatoire de la Côte d’Azur, CNRS, Artemis, Bd de l’Observatoire, F-06304 Nice, France\\
$^{5}$Department of Physics, University of Antwerpen, Prinsstraat 13, 2000 Antwerpen, Belgium\\
$^{6}$Department of Astrophysics/IMAPP, Radboud University, PO Box 9010, 6500 GL, Nijmegen, The Netherlands\\
$^{7}$Leiden Observatory, Leiden University, PO Box 9513, 2300 RA, Leiden, The Netherlands\\
}

\date{Accepted XXX. Received YYY; in original form ZZZ}

\pubyear{2024}

\begin{document}
\label{firstpage}
\pagerange{\pageref{firstpage}--\pageref{lastpage}}
\maketitle

\begin{abstract}
 
Galactic white dwarf binaries (WDBs) and black hole binaries (BHBs) will be gravitational wave (GW) sources for LISA. Their detection will provide insights into binary evolution and the evolution of our Galaxy through cosmic history. Here, we make predictions of the expected WDB and BHB population within our Galaxy. We combine predictions of the compact remnant binary populations expected by stellar evolution from the detailed Binary Population and Spectral Synthesis (BPASS) code, with a Milky Way analogue galaxy model from the Feedback In Realistic Environment (FIRE) simulations. We use \textsc{PhenomA} and \textsc{LEGWORK} to simulate LISA observations. Both packages make similar predictions that on average four Galactic BHBs and 673 Galactic WDBs are above the signal-to-noise ratio (SNR) threshold of 7 after a four-year mission. We compare these predictions to earlier results using the Binary Star Evolution (BSE) code with the same FIRE model galaxy. We find that BPASS predicts a few more LISA observable Galactic BHBs and a twentieth of the Galactic WDBs. The differences are due to the different physical assumptions that have gone into the binary evolution calculations. These results indicate that the expected population of compact binaries that LISA will detect depends very sensitively on the binary population synthesis models used and thus observations of the LISA population will provide tight constraints on our modelling of binary stars. Finally, from our synthetic populations we have created mock LISA signals that can be used to test and refine data processing methods of the eventual LISA observations.
\end{abstract}


\section{Introduction}

The study of gravitational wave (GW) signals in the mHz band will help researchers reveal characteristics of the processes of binary evolution and the history of our Galaxy. The observations of the future Laser Interferometer Space Antenna (LISA) mission will provide the first results in this observational window and will set the stage for a new era of astrophysics. Led by the European Space Agency (ESA), LISA is designed to detect GW signals in the frequency ranges from $10^{-4}$~Hz to 1~Hz. LISA is still in the development phase and is currently scheduled to launch in 2035. While the scientific community eagerly awaits its deployment, we have turned to simulations and predictive modelling to gain a glimpse of the signals LISA will eventually observe. By simulating data and predicting the GW signals that LISA is expected to detect, we can lay the groundwork for future discoveries and advance our understanding of the cosmos. In particular, LISA will be able to detect Galactic binaries and resolve many individual ultra-compact binaries \citep[UCBs, e.g., ][]{2023LRR....26....2A}. These resolved UCBs mostly consist of white dwarf binaries \citep[WDBs, e.g.,][]{2004MNRAS.349..181N, 2012ApJ...758..131N, 2017MNRAS.470.1894K, 2017ApJ...846...95K, 2019MNRAS.490.5888L, 2020ApJ...898...71B, 2020ApJ...898..133L, 2022MNRAS.511.5936K} and a smaller number of neutron stars and/or stellar black hole binaries \citep[BHBs, e.g.,][]{1990ApJ...360...75H, 2001A&A...375..890N, 2004MNRAS.349..181N, 2018MNRAS.480.2704L, 2020ApJ...892L...9A, 2020MNRAS.492.3061L, 2022ApJ...937..118W}.

Galactic WDBs have been recognised as the primary contributor to signals detected by space-based GW observatories \citep{1990ApJ...360...75H}.  However, there are three main source classes, namely the UCBs, massive binary black holes (MBHBs) and extreme mass ratio inspirals (EMRIs) as described within \citet{2023LRR....26....2A}. The majority of UCB sources are likely too distant to be individually detected by LISA. Instead they will contribute a significant stocastic GW signal in LISA's frequency band originating from the millions of unresolved WDBs systems in the Galaxy \citep{2003MNRAS.346.1197F,2009ApJ...705L.128R}. This is sometimes referred to as confusion noise. However, resolvable sources will stand out as distinct signals above the instrumental and confusion noise. It is expected there will be approximately $10^4$ resolved WDBs, as predicted by many studies such as \citet{2001A&A...365..491N}, \citet{2003MNRAS.346.1197F}, \citet{2010ApJ...717.1006R}, \citet{2017MNRAS.470.1894K} , \citet{2019MNRAS.490.5888L} and \citet{2020ApJ...898...71B}.

Although possible, detecting the stellar mass BHB population is a challenge; this is because they are rarer than WDBs and there are no known examples in our Galaxy as the only way to detect them will be with LISA. Until now, our knowledge comes from X-ray binaries and GW transients arising from the merger of BHBs \citep{2017hsn..book.1499C,2021arXiv211103606T}. The electromagnetically identified high-mass X-ray binaries that may eventually form BHBs in the Milky Way (MW) and nearby galaxies are also potential LISA sources \citep{2019IAUS..346....1V, 2012arXiv1208.2422B, 2013ApJ...764...96B}. While the BHB that cause GW transients would have been detectable by LISA before merging, if they had resided within our Galaxy.

Comprehending the Galactic sources of mHz GWs is essential to both understand the sources themselves but also to be able to then remove them to detect possible cosmological sources and other populations such as EMRIs \citep{2021PhRvD.103d4006C}. A binary population synthesis model is required to aid this search. In this work, we aim to use results from our own binary population synthesis results to predict the GW signals and understand what the signals are expected to be. At the same time we contrast and compare our predictions to the predictions of \citet{2018MNRAS.480.2704L, 2019MNRAS.490.5888L}, who used the same galaxy model, but a different binary population synthesis code. Thus, we can explore how predictions vary between population synthesis codes, and explore the reasons for differences between these predictions.

The work of \citet{2018MNRAS.480.2704L, 2019MNRAS.490.5888L}, presented a method for predicting the population of BHBs and WDBs in the MW using cosmological simulations with the Feedback In Realistic Environments\footnote{\url{https://fire.northwestern.edu/}} (FIRE) model \citep{2014MNRAS.445..581H}. These FIRE simulations deal with the formation and evolution of stars and galaxies over the history of the Universe. The simulations provide information on how many stars form through the age of the Universe and how their metallicity varies. They attached a binary population synthesis model to the FIRE results, creating synthetic DWD and BHB populations, by matching stellar population predictions to the FIRE simulation star formation, ages and metallicities.
 
In our study, we also use the same FIRE model to create a MW-like galaxy that is occupied by compact binaries, and then predict GW signals from these populations. Most binary population synthesis codes, such as the one that was implemented in \cite{2018MNRAS.480.2704L, 2019MNRAS.490.5888L}, are rapid population synthesis codes, which are known to be efficient. However, in our study, we use binary populations computed by the Binary Population and Spectral Synthesis (BPASS\footnote{\url{http://bpass.auckland.ac.nz} or \url{http://warwick.ac.uk/bpass}}) v2.2.1 code \citep{2017PASA...34...58E,2018MNRAS.479...75S, 2022ARA&A..60..455E}, which models the evolution of binary stars and stellar structure in detail rather than using analytic fits to single star models. This leads to different outcomes, especially increasing the stability of mass transfer \citep[see][]{2023MNRAS.520.5724B}. Which leads to differences in the predicted Galactic binary populations and the subsequent GW signals. By contrasting and comparing the predictions of the two sets synthetic GW signals from the BPASS and BSE binary populations we gain insights into the physics of interacting binaries.

This paper outlines the importance of detailed binary population synthesis codes for predicting GWs in LISA frequency band. In Section~\ref{sec:section2}, we describe how we create our Galactic binary populations, and examine the details of our simulated WDB and BBH populations. In Section~\ref{section:3}, we present the predicted GW signals for our synthetic populations and determine how many binary systems may be observed by LISA. In Section \ref{4}, we take the next step to calculate mock modulated GW signals from LISA for our different binary populations. Then in Section~\ref{section:5}, we discuss our results, comparing to other predictions by other studies, before presenting our conclusions in Section \ref{section:6}.

\section{FIRE and BPASS}
\label{sec:section2}

Our simulated predictions serve as crucial stepping stones, preparing researchers for the groundbreaking observations that LISA will provide once it is launched. This study combines a realistic model for the star formation history (SFH) and metallicity evolution for a Milky Way-like galaxy (Section~\ref{section:FIRE}), and a detailed binary population synthesis model, BPASS (Section~\ref{section:BPASS}).

\subsection{FIRE simulation}
\label{section:FIRE}

To be able to compare our work to the studies by \cite{2018MNRAS.480.2704L, 2019MNRAS.490.5888L}, we use the same MW-like galaxies from the FIRE suite \citep{2014MNRAS.445..581H}. We use the same m12i results, known as the `Latte' simulation \citep{2023ApJS..265...44W}. This simulation is based on the improved `FIRE-2' version of the code from \cite{2018MNRAS.480..800H} and ran with the code GIZMO\footnote{\url{http://www.tapir.caltech.edu/ phopkins/Site/GIZMO.html}} \citep{2015MNRAS.450...53H}. GIZMO solves the equations of hydrodynamics using the mesh-free Lagrangian Godunov `MFM' method. The analysis of the simulations is conducted using the publicly available Python package GIZMO ANALYSIS \citep{2020ascl.soft02015W}. For more detailed description of how FIRE was combined with the BSE binary population code see \cite{2018MNRAS.480.2704L, 2019MNRAS.490.5888L}.

\subsection{BPASS}
\label{section:BPASS}

BPASS is a suite of computer programmes that models the evolution of interacting binary stars. It  combines individual stellar models to create synthetic stellar populations and make varied predictions such as the spectral energy distributions of individual stars or of the total stellar population, the rates of electromagnetic and gravitational transients and much more \citep[see][and references therein]{2017PASA...34...58E, 2018MNRAS.479...75S, 2019MNRAS.482..870E,2022MNRAS.514.1315B}. The stellar evolution models incorporate the physical processes such as mass transfer (MT), stellar winds, and supernova explosions. The main unique feature of BPASS \citep[first described in][]{2008MNRAS.384.1109E} is that the stellar models are evolved in detail using a custom version of the Cambridge STARS code \citep{1971MNRAS.151..351E}. This allows the reaction of the stellar structure to binary interactions to be evolved in detail. This typically lead to MT that is more stable in our models than is commonly assumed in other population synthesis codes \citep{2023MNRAS.520.5724B}. Another feature of BPASS is the predictions have been widely tested against a large range of observations. This list is growing all the time, for example see \cite{2023ApJ...943L..12K} who study the impact of binary interactions on chemical enrichment of galaxies using BPASS and several other binary population synthesis code results. In this work we aim to take steps to make BPASS predictions on future LISA observations of WDBs and BHBs in our Galaxy.

Many other stellar population synthesis codes are rapid binary population synthesis codes \citep[e.g.][]{2002MNRAS.329..897H,2020ApJ...898...71B,2022JOSS....7.3838C}. The investigations by \citet{2018MNRAS.480.2704L, 2019MNRAS.490.5888L} used the binary stellar evolution (BSE) code \citep{2002MNRAS.329..897H} where the massive binary star treatment had been modified. BSE follows the evolution of binary stars with the evolution approximated using analytic fits to single star models from \citet{2000MNRAS.315..543H} and the binary evolution described in \cite{2002MNRAS.329..897H}. This approach allows for very computationally efficient exploration of a wide range of parameters. BSE models, however, miss important effects in the stellar structure and MT compared with detailed binary population synthesis codes \citep[e.g.,][]{2021ApJ...922..110G, 2021A&A...645A..54K, 2021A&A...650A.107M, 2023MNRAS.520.5724B}. Detailed structure evolution calculations result in stable MT (SMT) being more common than found by rapid codes. SMT occurs when matter is transferred from one star to its companion without the orbit dramatically shrinking and the binary entering a phase of common-envelope evolution. SMT significantly influences binary stars evolution and is important for the formation of many types of binary compact objects. However, SMT mechanisms are complex and uncertain, with factors like accretion rates, angular momentum transfer, and interaction between stellar envelopes playing important roles. \citet{2023MNRAS.520.5724B} and \citet{2023MNRAS.523.1711G} discussed the effects of SMT in BPASS models, and postulated that SMT with super-Eddington accretion is responsible for high mass BH population with masses 35-100$\rm M_\odot$ as observed by the LIGO-VIRGO-KAGRA collaboration.

In this work, we use the stellar models from BPASS v2.2.1. The population of WDBs and BHBs has been calculated using the GW population synthesis code \textsc{Tui} (v2.jje), a GW binary population synthesis programme which is described in \citet{2022MNRAS.511.1201G}, \citet{2023NatAs...7..444S}, \citet{2023MNRAS.520.5724B} and \citet{2023MNRAS.524.2836V}. The predictions are made using the fiducial BPASS initial binary parameters and initial mass functions. The initial mass function is model {135--300} described in table 1 in \citet{2018MNRAS.479...75S}, which is the IMF from \citet{1993MNRAS.262..545K} with an upper mass limit of 300~M$_{\odot}$. We use the initial binary parameter distribution of \citet{2017ApJS..230...15M}. The kick used in the population synthesis is the Hobbs neutron star kick, which is adapted to a momentum kick for BHs \citep{2005MNRAS.360..974H}. \textsc{Tui} takes the final binary parameters after both stars have a stellar remnant. It then evolves the binary using the equations of \citet{1964PhRv..136.1224P} to model the impact of gravitational radiation removing energy from the orbit. Unless the systems merge we stop integration of the orbit at 100~Gyrs. The resultant model population is binned by age, frequency and remnant masses at a resolution of 0.1 dex, while binary eccentricity has bin sizes of 0.1.

To combine the binary population with the FIRE galaxy we build on our work in \cite{2023MNRAS.524.2836V} where we used these same \textsc{Tui} models to model individual stellar clusters. \textsc{Tui} outputs the numbers of WDBs and BHBs expected, with different masses, periods, eccentricities at different ages of the stellar population, per M$_{\odot}$ of star formation at a number of initial metallicities. This can then be combined with the FIRE simulation. The FIRE simulation is a list of ``star'' particles each with an initial mass, age and metallicity. We assume that each particle is equivalent to a star cluster. Thus the mass, age and metallicity are used to identify a parameter space in the \textsc{Tui} outputs to then estimate how many binary compact objects each star particle contains. For each star particle we simply multiply the BPASS-predicted WDB and BHB numbers per stellar mass (which can be a fraction) by the initial star particle mass. Each integer value of binaries is then kept, the remainder is carried over to the next star particle rather than being lost. It is important to note that we work through the list of star particle in age order. Therefore carried over values are associated with systems with similar age, thus the distribution of the eventual binary systems within the synthetic catalogue will be similar. This method means we do not underestimate the total binary population; however, it might mean we miss rare types of binary systems. The fraction of stars carried over is typically small, of the order of 0.1 for WDB or of the order of $10^{-8}$ for BHB. Another way to evaluate the importance of the carry over is that there are 13,976,485 star particles in the galaxy model we use. This means for the full frequency range of binaries, on average for BPASS each star particle hosts 19.52 WDBs and 0.35 BHBs. Given the particles are have the same age suprious errors from carrying over the IMF weighting should be minor.

The final step is to randomly distribute the binary within the \textsc{Tui} output bin using a uniform random distribution. This avoids all the binaries having exactly the same parameters. This process results is a list of compact remnant binaries and their distribution within the synthetic galaxy. This allows for direct comparison to the predictions of \cite{2018MNRAS.480.2704L, 2019MNRAS.490.5888L} who used the same synthetic galaxy but a different binary population synthesis code and different method of distributing star masses; the different ways of distributing star masses may not result in significant differences in mass distributions in the galaxy; however, this needs to be verified. The aim of our study is to provide insight into the importance of how the binaries are modelled for predicting the LISA observable population.

\subsection{Comparing the BSE and BPASS synthetic populations in FIRE galaxy m12i}
\label{section:FIREBPASS}
Having matched the FIRE m12i star particles with the binary population synthesised by BPASS, we can now compare to the BSE models of \citet{2018MNRAS.480.2704L, 2019MNRAS.490.5888L}. Both sets of calculated outputs include Cartesian galactic coordinates, orbital frequency, primary and secondary masses, age and metallicity of each binary. These outputs are used to create synthetic LISA observations in Sections~\ref{section:3} and \ref{4}.

\subsubsection{The number of Galactic WDB and BHB}

\begin{figure*}
    \centering
    \includegraphics[width=2\columnwidth]{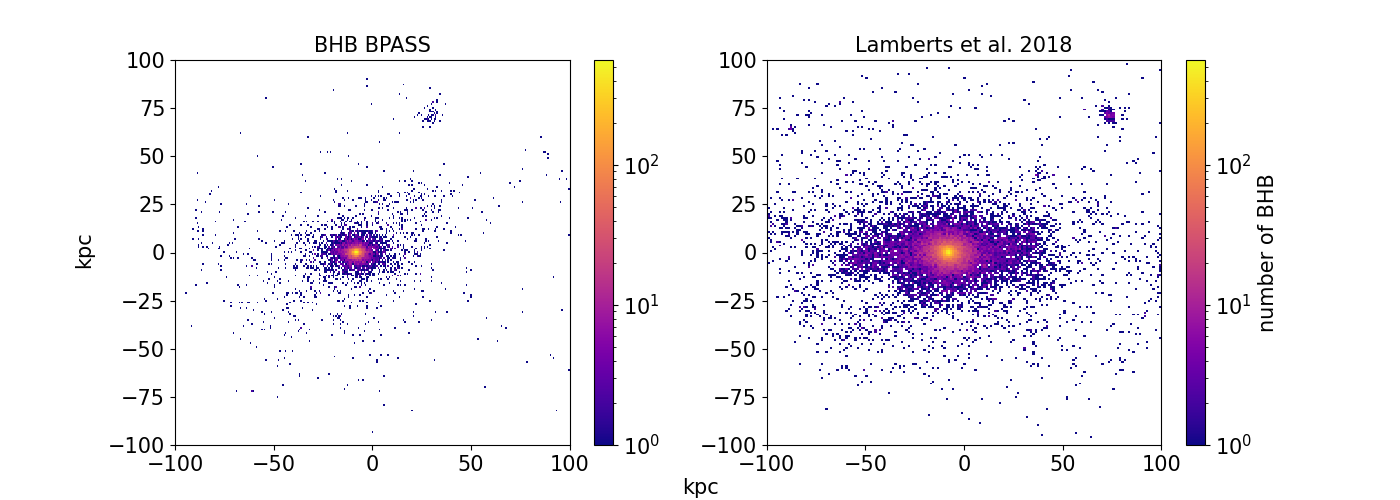}
    \caption{ Large-scale 2D number density of BHB populations in the simulated galaxy from BPASS and BSE model populations viewed face on. These are the BHBs that are within the with frequencies $>10^{-5}$~Hz. BHBs are present in the galactic bulge, disks, galactic halo and a satellite galaxy in both model populations. The population density is the also highest at the centre of the galaxy for both models.}
    \label{fig:number_bhbh}
\end{figure*}

\begin{figure*}
    \centering
    \includegraphics[width=2\columnwidth]{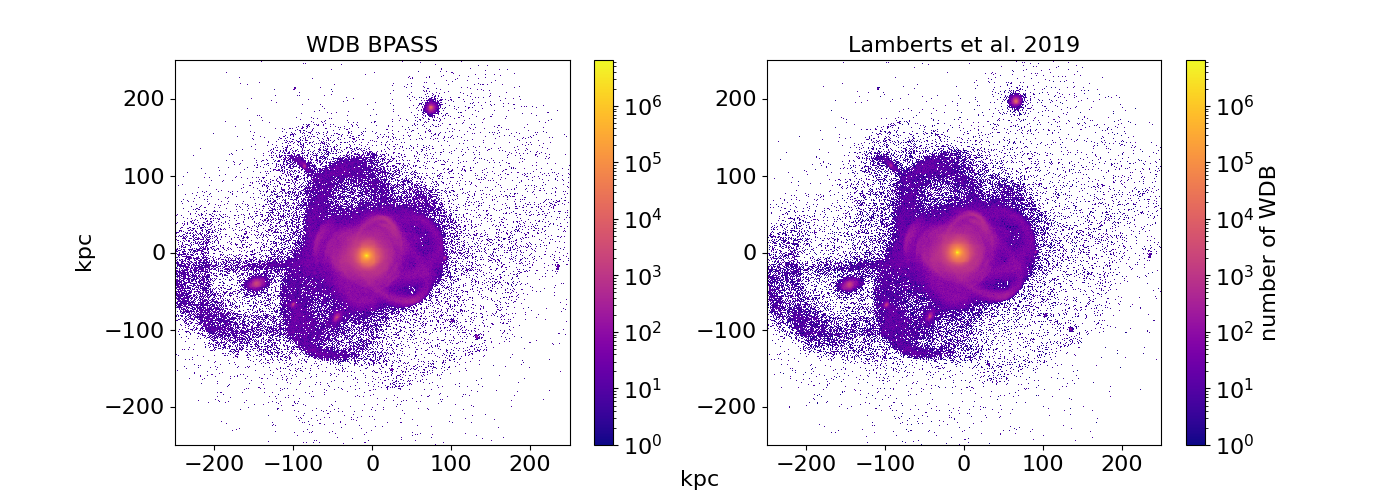}
    \caption{ Large-scale 2D number density of WDB populations in the simulated galaxy from BPASS and BSE model predictions viewed face on. WDBs are present in the galactic bulge, disks, galactic halo, tidal streams and satellite galaxies in both model predictions. The population density is the highest at the centre of both models.}
    \label{fig:number_wdwd}
\end{figure*}

\begin{table}
  \centering
  \caption{ Number of Galactic binary systems in the MW-like galaxy predicted by two models in the frequency range of $10^{-5}$Hz to 1 Hz. The full predicted population from BPASS of frequency ranging from $10^{-10}$ Hz to 1 Hz is given in parentheses.}
    \begin{tabular}{r|rr}
           Events & BSE & BPASS \\
          \hline
    BHB & 25,335 & 9,298\\ 
     &  & (495,586) \\ 
    
    WDB & 35,832,029 & 58,522,007   \\
     & & (272,789,350)   \\
    \end{tabular}%
  \label{tab:numbersource}%
\end{table}%
The numbers of Galactic binary systems predicted by the two binary population synthesis models are shown in Table~\ref{tab:numbersource}. There are fewer BHB and more WDB systems predicted by BPASS than \cite{2018MNRAS.480.2704L, 2019MNRAS.490.5888L} respectively. However, the number of WDB systems in both samples is closer in magnitude than the BHB populations. Most of these binaries are expected to be less than 20 kpc from the Galactic center \citep[c.f][]{2004MNRAS.349..181N, 2019MNRAS.490.5888L, 2002MNRAS.333..469S}. However because of the Galactic halo and tidal streams, some objects are at greater distances of up to 300 kpc, as shown in Figures \ref{fig:number_bhbh} and \ref{fig:number_wdwd}. 

\begin{figure}
    \centering
    \includegraphics[width=\columnwidth]{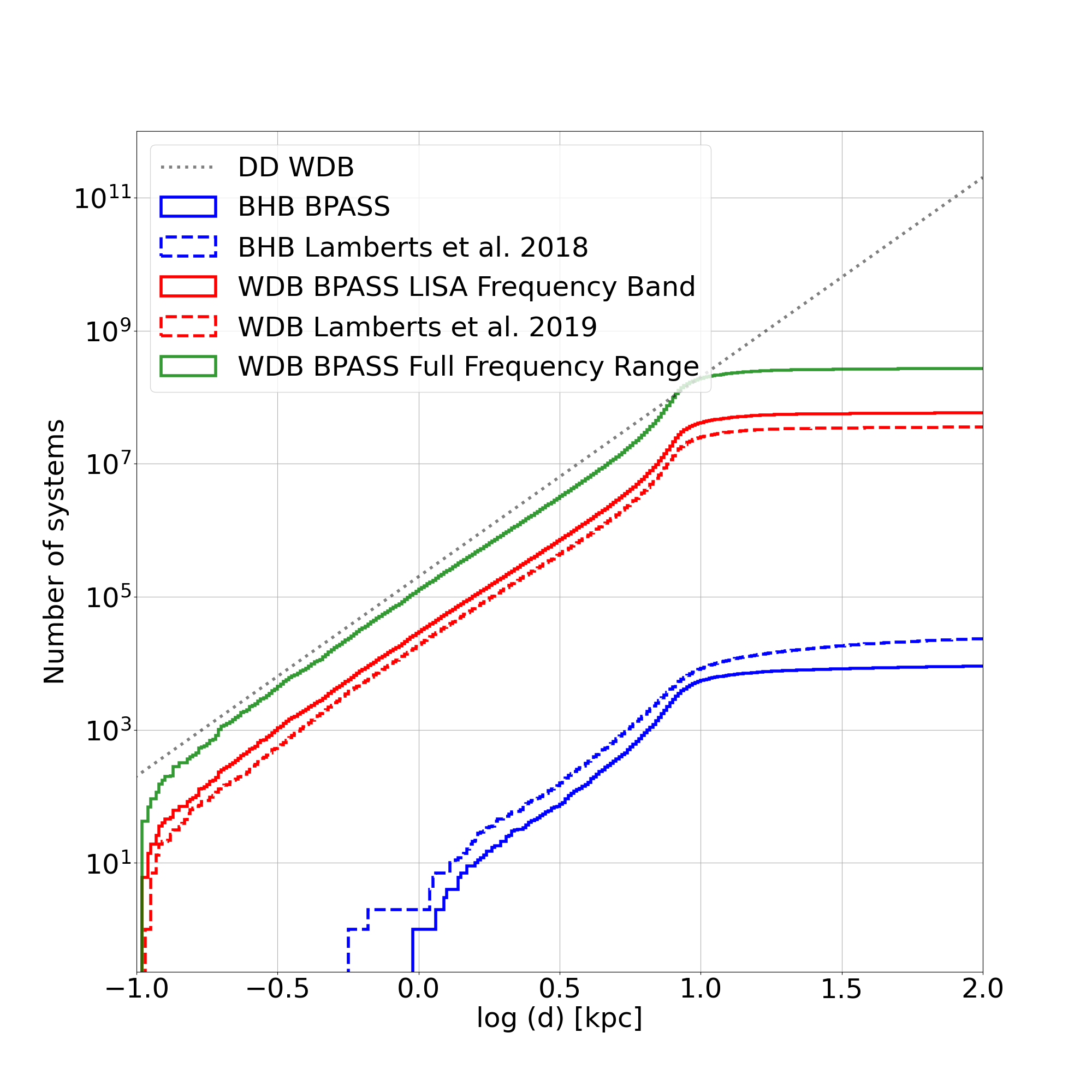}
    \caption{Cumulative number distribution of binary systems at different distances from the Sun. Blue and coloured lines represent Galactic BHB and WDB populations with frequencies above $10^{-5}$~Hz, respectively. The BPASS models are presented by solid lines and the BSE models by dashed lines. The solid green line represents BPASS WDBs with full frequency range. The observationally driven extrapolated DDWDBs population prediction is represented by the dotted straight line.}
    \label{fig:number_distance}
\end{figure}

\subsubsection{Spatial density of Galactic WDBs}

We have attempted to verify if our WDB model reproduces the expected space density of WDBs around the Sun. In Figure~\ref{fig:number_distance} we show the expected number of WDBs and BHBs as we increase the distance of the objects included with the observed volume. The dotted line represents the Double Degenerate WDs (DDWDs) space density prediction, that are extrapolated from the DDWD population within 25 pc of the Sun surveyed by \cite{2016MNRAS.462.2295H}. Here we use these DDWDs to verify the space density of the modeled WDBs. The space density of the WD population is estimated to be $4.8 \pm 0.5 \times 10^{-3} \mathrm{pc}^{-3}$ at the Sun's location. Within this WD population, 26 per cent are binaries, and 10 per cent are DDWDs. Using this estimated observed space density we can extrapolate how many WDBs should be within the Galaxy. In Figure~\ref{fig:number_distance} we can see that the BPASS WDB predictions are closer to the number of DDWDs from the extrapolation of the local space density. While BSE is an order of magnitude below the extrapolation. Our BPASS predictions includes the full frequency range of orbital periods ($10^{-10}$ - $1$ Hz), this is the main reason in the difference between BPASS and BSE and why the former matches the extrapolated DDWD space density within 10 kpc. We note that this extrapolation of the observed space density is approximate and does not take into account the Galactic disk structure, and the turnover in the theoretical numbers beyond 10~kpc demonstrates this within Figure~\ref{fig:number_distance}.

\subsubsection{The total binary mass distribution}

A summary of the distribution of total binary masses for the synthetic binary populations is shown in Table~\ref{tab:totalmass}. The total mass of the BSE BHB population has higher median and mean values compared with the BPASS BHB population; and the maximum BHB in BPASS is more then twice the maximum BHB mass from BSE models. The total binary mass of BPASS WDB population has a lower mean and higher median values than that of the BSE; more low mass WDBs survive in BPASS than in BSE.
\begin{table}
  \centering
  \caption{ Summary of the mean, median, minimum, maximum and standard deviation values of the total binary mass distribution of each WDB and BHB population with frequencies between $10^{-5}$Hz and 1 Hz. The full predicted population from BPASS of frequency ranging from $10^{-10}$Hz to 1 Hz is given in parentheses.}
  \begin{tabular}{lcccc}
   & \multicolumn{2}{c}{BHB} & \multicolumn{2}{c}{WDB} \\
   & BSE & BPASS & BSE & BPASS \\
\hline
    Maximum & 85.40$M_\odot$ & 191.10$M_\odot$ & 2.39$M_\odot$ & 2.38$M_\odot$ \\
            &    &   (249.60$M_\odot$) &    &  (2.82$M_\odot$)  \\
    Median & 25.53$M_\odot$ & 19.91$M_\odot$ & 0.90$M_\odot$ & 1.03$M_\odot$ \\
            &     &  (25.98$M_\odot$) &   &  (1.02$M_\odot$)  \\
    Std. dev. & 13.3$M_\odot$ & 19.4$M_\odot$ & 0.34$M_\odot$ &  0.29$M_\odot$ \\
     &       & (20.81$M_\odot$)  &   & (0.27$M_\odot$)  \\
    Mean & 28.17$M_\odot$ & 26.21$M_\odot$ & 1.03$M_\odot$ & 1.07$M_\odot$ \\
            &    &   (30.10$M_\odot$) &    &  (1.07$M_\odot$)  \\
    Minimum & 6.72$M_\odot$ & 5.18$M_\odot$ & 0.39$M_\odot$ & 0.32$M_\odot$ \\
            &    &   (4.50$M_\odot$) &    &  (0.22$M_\odot$)  \\ 
  \end{tabular}
  \label{tab:totalmass}
\end{table}

\begin{figure*}%
    \centering
    \includegraphics[width=2\columnwidth]{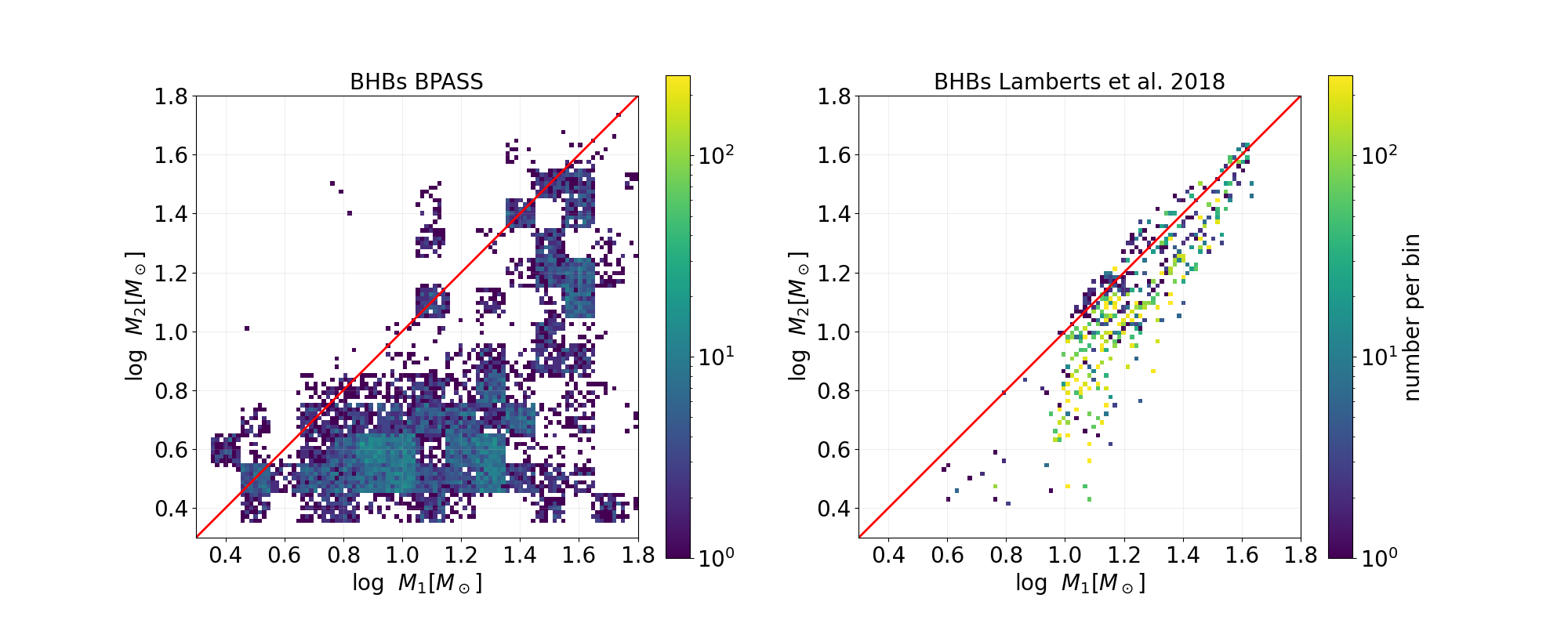} 
    \caption{ The distribution of primary $\rm M_1$ and secondary $\rm M_2$ masses in BHBs from BPASS (left) and BSE (right). The red line represents $\rm M_1=M_2$, where above the line $\rm M_1<M_2$ and below the line $\rm M_1>M_2$. }
    \label{fig:bhbh_m12}
\end{figure*}

\begin{figure*}%
    \centering
    \includegraphics[width=2\columnwidth]{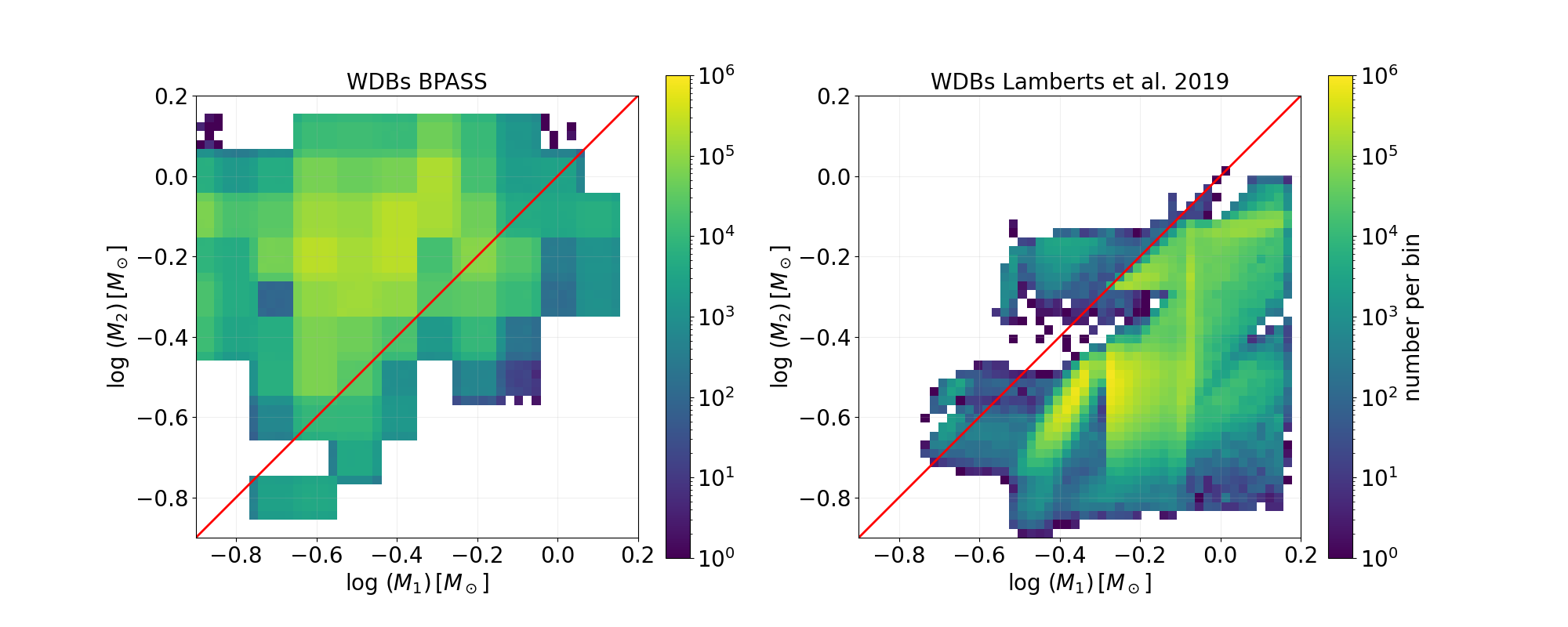} 
    \caption{  The distribution of primary $\rm M_1$ and secondary $\rm M_2$ masses in WDBs from BPASS (left) and BSE (right). The red line represents $\rm M_1=M_2$, where above the line $\rm M_1<M_2$ and below the line $\rm M_1>M_2$. }
    \label{fig:wdwd_m12}
\end{figure*}

The BPASS and BSE codes have differences in their assumed binary evolution models. This leads to different evolutionary outcomes for stars of the same initial mass, which here can be seen in the predicted masses of the WDB systems. In this we consider the primary star to always be the initially more massive star. MT in a binary occurs at different points of evolution, determined by the binary physics included as well as the initial masses and separations of the stars. While some of the differences are shown in Table~\ref{tab:totalmass} they are also apparent in the distribution of the masses of each remnant in a binary. We show in Figures~\ref{fig:bhbh_m12} and~\ref{fig:wdwd_m12} the distribution of the masses for both remnants in the of BHBs and WDBs respectively. For the BHBs we see a wider range of BHB masses, as well as examples of the most massive BH coming from the initially less massive star in BPASS results. This is also apparent for the WDBs where for BPASS the majority of the most massive WDs are from the initially less mass star. 

For the BHBs, we expect the broader range of BH masses for BPASS is inherent in the stellar evolution models. If we compare the initial-to-final mass relations from the codes, using figure 2 of \citet{2018MNRAS.480.2704L} and figure 17 of \citet{2017PASA...34...58E}, the range of black hole masses from a certain initial mass is wider for BPASS than BSE. This is likely due to the detailed stellar-evolution models that are calculated for BPASS populations. The mass loss in binary interactions has a more significant impact on core masses and thus remnant masses than found in rapid codes. The impact of binary interactions on the remnant masses in detailed stellar evolution models has also been investigated by \citet{2021A&A...656A..58L} and \citet{2022MNRAS.511..903P}.

For the BHBs this mass reversal is relatively rarer than for WDBs. We find this is because super-Eddington accretion onto the first BH is formed, it remains the most massive BH for most binaries as discussed by \citet{2023MNRAS.520.5724B}. While for the WDBs we find in BPASS that MT for low mass stars is more stable than for BSE with more mass exchange from the primary to secondary star with fewer occurrences of common-envelope evolution (CEE). Especially during the first occurrence of Roche Lobe Overflow (RLO) before the primary star becomes a WD. The reason for this stability is that BPASS uses detailed stellar evolution models for the donor star so MT stability is determined by how the star's radius responds to mass loss. If the mass loss stops the radial expansion of the donor than mass transfer is taken to be stable. While if it continues to grow then a CE phase occurs. For BSE, as it is a rapid code, limits on stability are assumed determined by the evolutionary state of star and the mass ratio at onset of MT \citep[e.g.][]{2002MNRAS.329..897H}. This may overestimate the number of systems that experience unstable MT. Such increased MT stability has been found by others who have used detailed stellar evolution models to follow binary interactions \citep[e.g][]{2021ApJ...922..110G, 2021A&A...645A..54K, 2021A&A...650A.107M, 2023MNRAS.520.5724B}.This higher occurrence of MT stability is the leading cause of the differences in predicted GW signals for WDBs between BPASS and BSE. We find that 93 per cent of BPASS WDBs in the LISA band have the most massive WD from the secondary star, compared to only 1.2 per cent for the BSE models. The difference is less extreme for the BHBs with only 7.0 per cent of BPASS secondary stars being the most massive BHs, with 4.5 per cent for the BSE model.

\subsubsection{Galactic map of Galactic WDBs and BHBs}
Our next step is to understand whether LISA will be able to observe these WDBs and BHBs. To this end we assume that the LISA constellation's position is at the Sun's location, at a distance of 8~kpc from the model galaxy's center. Thus the distance to each of the sources is computed from this point. The large-scale density maps of the Galactic binary population densities are shown in Figures~\ref{fig:number_bhbh} and~\ref{fig:number_wdwd}, where all colour bars represent numbers of binary systems in each bin. The inner 30 kpc for the Galactic centre are presented in appendix Figures~\ref{fig:number_bhbh_short} and~\ref{fig:number_wdwd_short}.

On the large scale map of 100 kpc radius about the LISA constellation shown in Figure~\ref{fig:number_bhbh}, both galaxies contain a visible satellite galaxy. On a small-scale, shown in Figure~\ref{fig:number_bhbh_short}, around a 25~kpc radius about the Sun's position, there are noticeable differences between the two BHB galaxies. BPASS BHBs are found mostly near the bulge, and \cite{2018MNRAS.480.2704L} BHBs spread out more evenly from the bulge towards the edge of the Galactic disk and halo, this can be seen in Figure~\ref{fig:number_distance} as well. The distributions over the in the inner 10~kpc of the Galaxy are similar but with BSE predicting many more BHBs than BPASS. However, beyond 10~kpc the difference between BSE and BPASS becomes greater. With BSE predicting many more BHBs beyond this distance, as can be seen in Figure \ref{fig:number_distance}. This difference is because we begin to sample the galactic halo where there are more low metallicity stars so the difference between BSE and BPASS in this regime become apparent.

Figure~\ref{fig:number_wdwd} (small-scale map in appendix~\ref{fig:number_wdwd_short}) shows the distribution of WDB systems. Both galaxies appear to be similar in structure. Furthermore, both galaxies contain visible satellite galaxies and tidal streams. These satellite galaxies' sizes are similar to the Small Magellanic Cloud in the MW galaxy.

\begin{figure*}
    \centering
    \includegraphics[width=2\columnwidth]{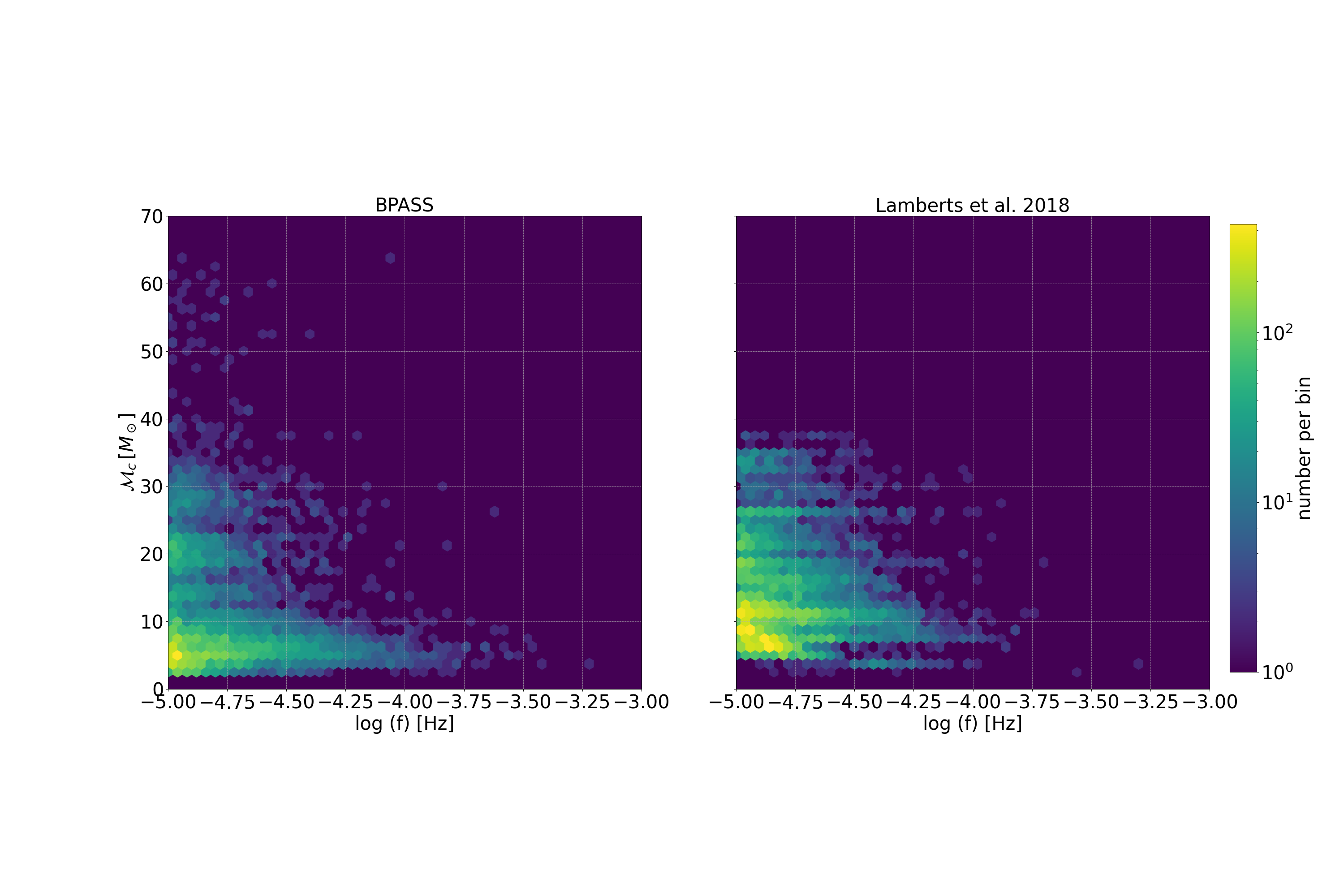}
    \vspace*{-15mm}
    \caption{ Distribution of BHBs over $\mathcal{M}_c$ and frequency, BPASS (left) and BSE (right). The BPASS BHB population has more systems with higher $\mathcal{M}_c$ than BSE.}
    \label{fig:bhbh_fMc}
\end{figure*}

\begin{figure*}
    \centering
    \includegraphics[width=2\columnwidth]{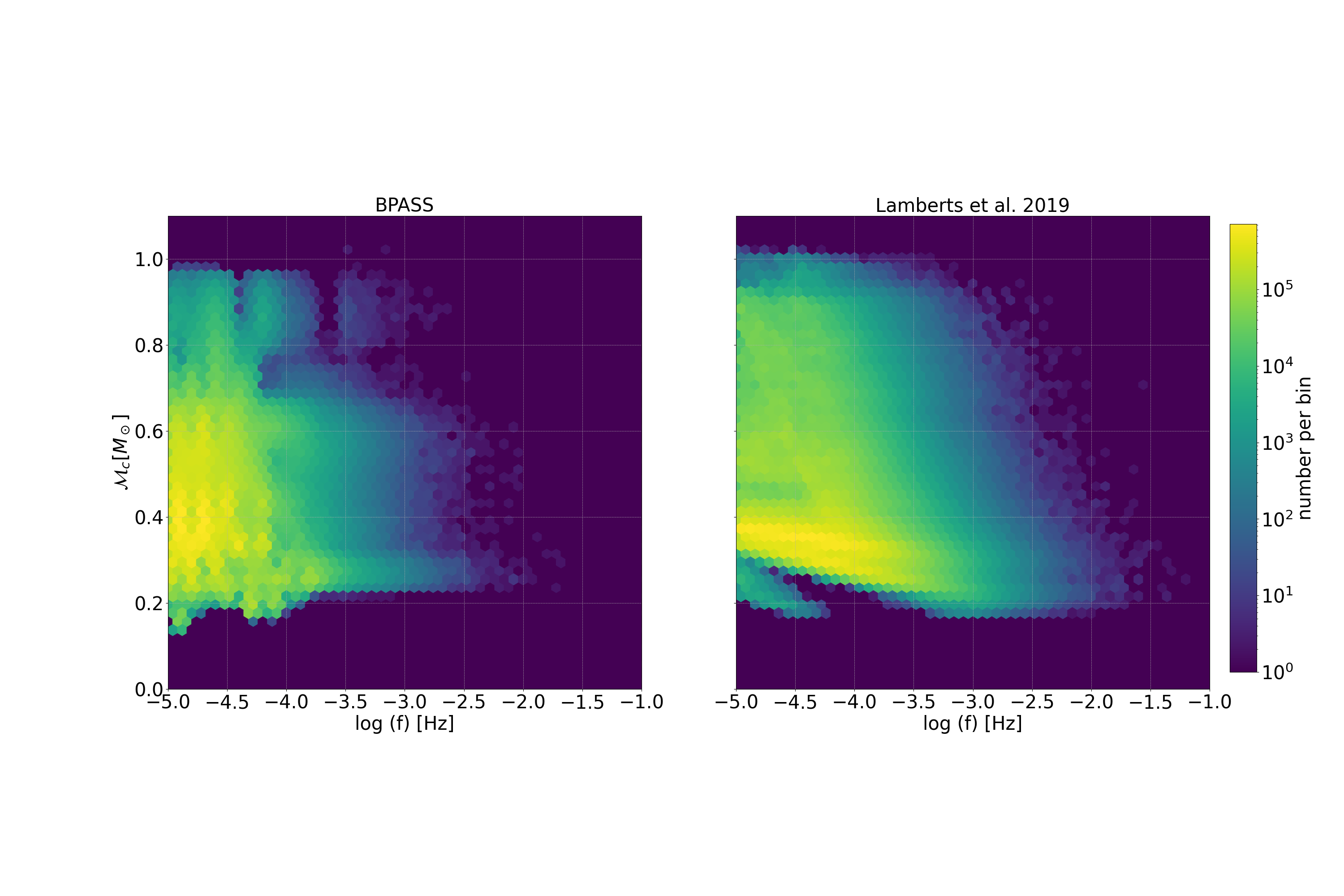}
    \vspace*{-15mm}
    \caption{ Distribution of WDBs over $\mathcal{M}_c$ and frequency, BPASS (left) and BSE (right). There are clear differences in the distributions between the two models that likely result from the different treatment of mass transfer and modelling of stellar structure.}
    \label{fig:wdwd_fMc}
\end{figure*}

\subsubsection{$\mathcal{M}_c$ distribution}
In GW observations a key parameter for any binary is the chirp mass $\mathcal{M}_c$. It is a combination of the component mass and is defined as
\begin{equation}
     \mathcal{M}_c=\frac{(M_1 M_2)^{3/5}}{(M_1 + M_2)^{1/ 5}}
\end{equation}
where $\rm M_1$ and $\rm M_2$ are primary and secondary masses of the binary respectively. The amplitude of GWs is related to the frequency and $\mathcal{M}_c$ of the binary source. Higher $\mathcal{M}_c$ generally correspond to more massive BHs and WDs and stronger GW signals. We find that BPASS predictions for BHB masses are higher, shown in Table \ref{tab:totalmass}. While in Figure \ref{fig:bhbh_fMc} we can see that only BPASS results have BHB with $\mathcal{M}_c$ above 40~M$_{\odot}$. One reason for this is related to the initial mass ranges of stars used. In \citet{2018MNRAS.480.2704L} the initial mass of the stars has a maximum of 120~M$_{\odot}$ with the assumption more massive stars all experienced Pair-Instability Supernova (PISNe) and so produce no black holes. However, in BPASS the maximum initial stellar mass used is 300~M$_{\odot}$ and PISNe are assumed to occur when the star ends its evolution with a helium-core mass in the range between 64 and 133~M~$_{\odot}$ thus preventing black holes with high masses from forming. As we note above, binary interactions in detailed evolution models have a more significant impact on the core masses and thus eventual fate of stars. Thus while BHs from stars above 120~M$_{\odot}$ are rare they do occur. A second, and the more significant reason, is the super-Eddington accretion assumption within BPASS populates this upper range of masses as detailed in \citet{2023MNRAS.520.5724B} \citep[see also][]{2020ApJ...897..100V, 2023MNRAS.523.1711G}. Thus these differences together explain the significant difference in the maximum chirp masses of the BHB predictions.

Next if we look at chirp masses below 40~M$_{\odot}$ shown in Figure~\ref{fig:bhbh_fMc}, the general shape of the BPASS and BSE distributions are similar. With higher frequencies for lower mass binaries. With the majority of BHBs being in a $\mathcal{M}_c$ range from 3~$M_\odot$ to 10~$M_\odot$. The BSE BHB population is more numerous than the BPASS BHB population, and has multiple systems with low $\mathcal{M}_c$ and low frequencies. There are also more BPASS BHB systems present at the higher end of the LISA frequency band where LISA is most sensitive ($\rm 10^{-3}~Hz$).

As discussed above the stability of MT in interacting binary stars is a key driver of the WDB mass distribution, affecting both the masses and the period distributions which are important for predicting the LISA signal from these objects. We can see the impact in the different distributions for BPASS and BSE in Figure~\ref{fig:wdwd_fMc}. There is a large triangular gap in the BSE distribution from $\log(\mathcal{M}_c/M_{\odot})$=0.3 to 0.2 at frequencies of $10^{-4.5}$ to 10$^{-3}~{\rm Hz}$. While in the BPASS results this region is populated. This is the result of the more stable model of MT in BPASS leading to more extremely low-mass WDs allowing for the broader range of possible WDB chirp masses. We find that distribution of chirp masses varies with age of the WDBs (see Figure~\ref{fig:age}). With lower mass chirp masses being more likely for older WDBs. Overall, there are fewer WDBs within the LISA frequency range. This is another result of the BPASS MT stability, because more stable MT means fewer CEE systems with their shorter periods. In addition the BPASS CEE prescription is relatively weaker than these frequently used in BSE codes \citep[see discussion in ][]{2023NatAs...7..444S}. Both of these mean that while we have more WDBs (because of fewer mergers in CEE) we have fewer WDBs with orbits in the observable range.

Finally, we also note that one factor that also changes close binary periods is magnetic-wind braking \citep[e.g][]{1983ApJ...275..713R} for systems involving a single WD. As BPASS does not include this effect, in could be another reason why BPASS has fewer short period WDBs compared to BSE.

\begin{figure}%
    \centering
    \includegraphics[width=\columnwidth]{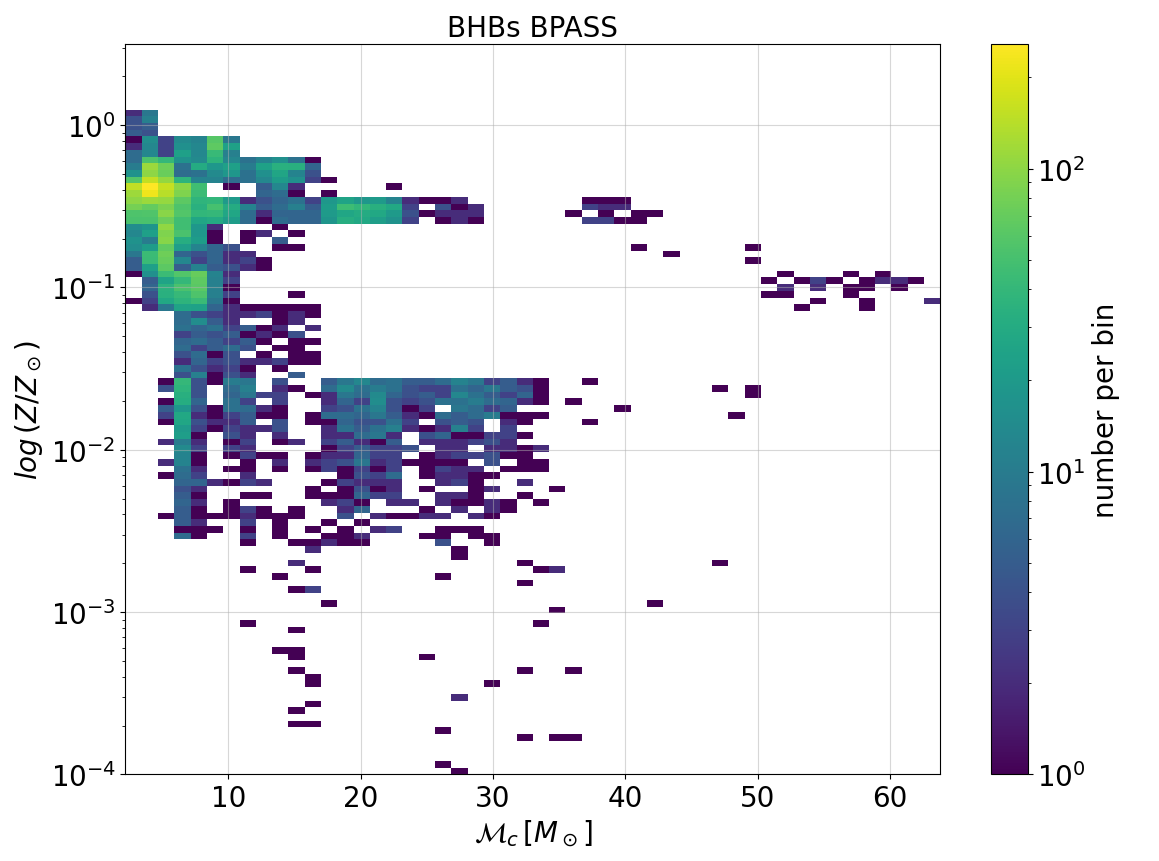} %
    \caption{ The distribution of BPASS BHBs over $\mathcal{M}_c$ and metallicity of the progenitor stars.}%
    \label{fig:bh_Z}%
\end{figure}

\begin{figure*} %
    \centering
    \begin{minipage}[b]{0.5\textwidth}
        \centering
        \includegraphics[width=\textwidth, height=0.76\textwidth]{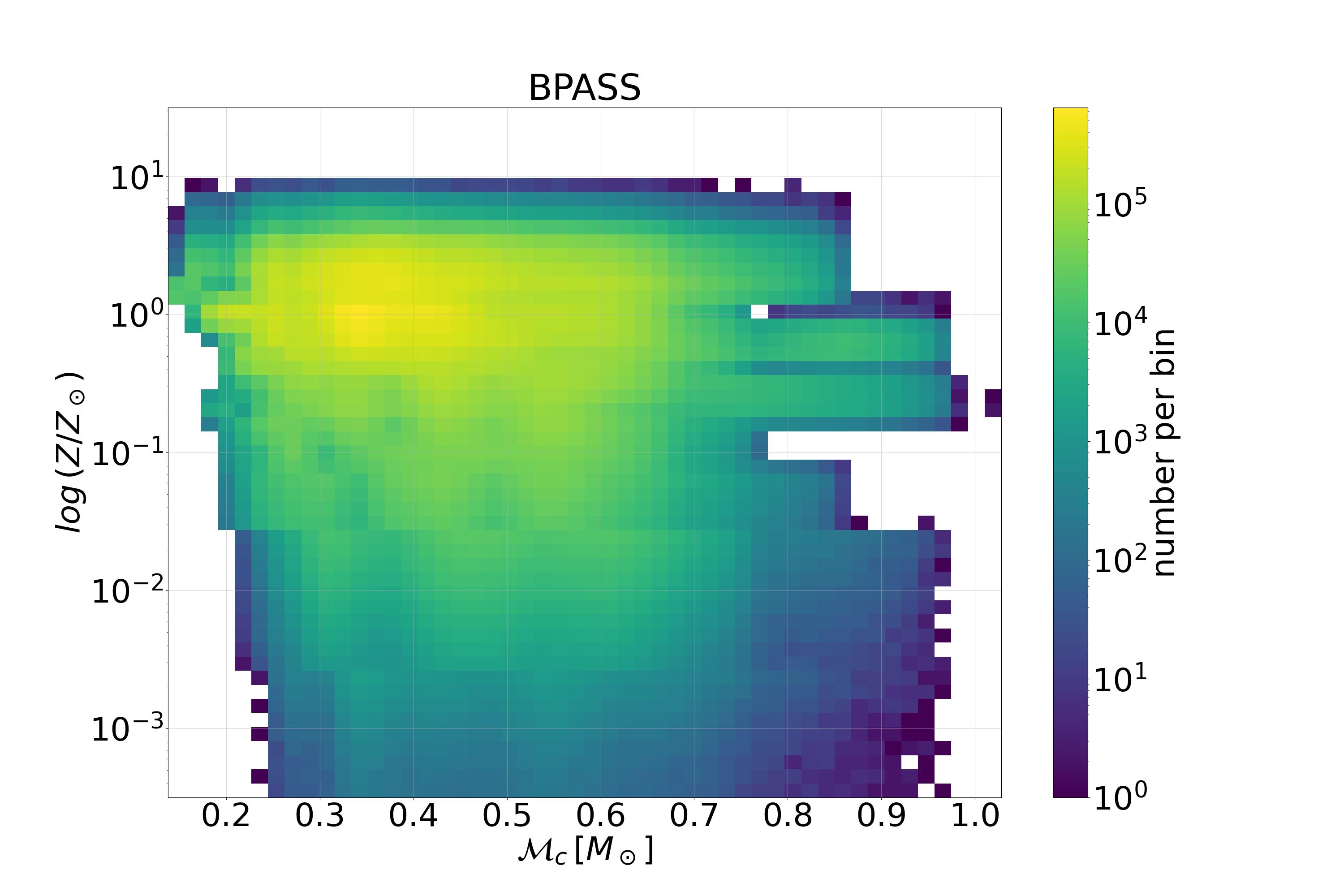}
    \end{minipage}\hfill
    \begin{minipage}[b]{0.5\textwidth}
        \centering
        \includegraphics[width=\textwidth, height=0.76\textwidth]{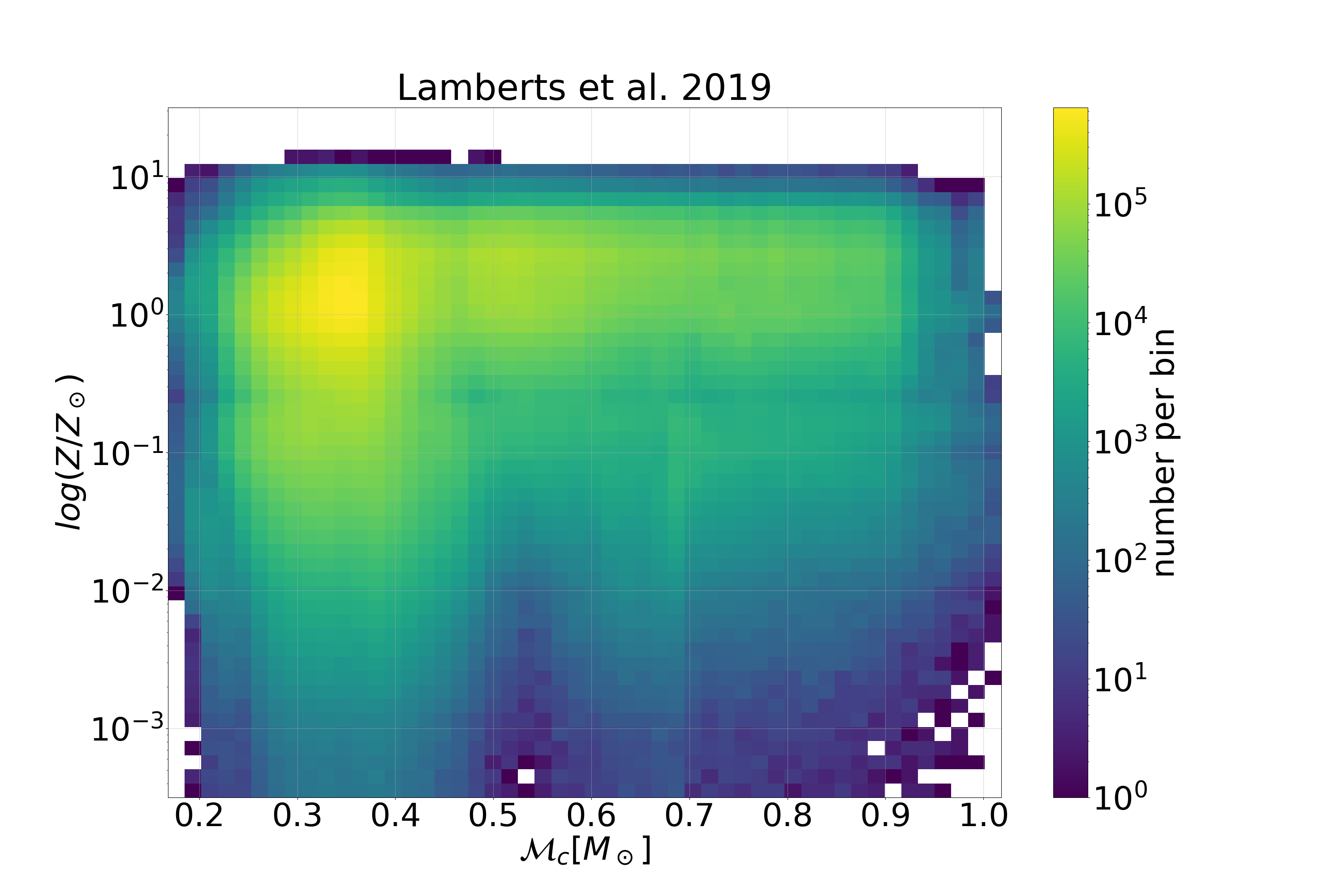}
    \end{minipage}
    \caption{ The distribution of BPASS (left) and BSE (right) WDBs over $\mathcal{M}_c$ and metallicity of the progenitor stars.}
    \label{fig:wd_Z}
\end{figure*}

\subsubsection{Metallicity distribution}

The metallicity of a binary system affects its evolution and GW emission in several ways. Lower metallicity stars have weaker stellar winds and so can form more massive remnants at the end of their lives. We see this in Figure~\ref{fig:bh_Z} for the BPASS results, the BHB have higher masses at lower metallicities. Also with weaker winds stars can grow to larger radii so that binary interactions can become more common. Conversely this means that it is possible binary interactions are less likely at high metallicity  as the envelope will be removed before it can interact. Most of the BHB systems are found below solar metallicity $Z_\odot$ = 0.02. BHB systems with $\mathcal{M}_c$ greater than 50 M$_\odot$ only exist in a narrow metallicity band of around one-tenth solar metallicity. These most massive BHB systems above 50~M$_{\odot}$ are the result of SMT during the evolution of both the primary and secondary, with super-Eddington accretion onto the BHs leading to the most massive BHs. 

We note that below a metallicity of $Z=0.004$, we allow secondary stars in a binary to experience quasi-chemically homogeneous evolution. This is an atypical evolutionary pathway that is the result of mass transfer onto massive stars making the rotate so rapidly that rotationally induced mixing means they are fully-mixed stars during their main-sequence evolution \citep{2005A&A...443..643Y, 2007A&A...465L..29C}. This is assumed for stars at low Z ($Z \leq 0.004$), that are initially more massive than 20~M$_{\odot}$ and they accrete more than 5 per cent of their initial mass from the primary star. These stars are assumed to rapidly rotate and become fully mixed during their main sequence evolution. As such they never fill their Roche lobe and donate material to the primary BH so it remains at its birth mass \citep[see][for further details]{2023MNRAS.520.5724B}. The implication here is that for such systems SMT onto the primary BH does not occur. This is the reason behind the horizontal linear feature at a $\log(Z/Z_{\odot})=10^{-0.5}$. Up to this metallicity QHE does not occur, but below this where QHE is assumed to occur, SMT onto BHs only occurs for a few systems. Below this metallicity we see a much more gradual increase in the masses of black holes as metallicity decreases. This is similar to what we expect occurs in BSE. While we do not have the equivalent data for BSE, Figure 2 in \citet{2018MNRAS.480.2704L} allows us to infer that BSE has a regular increase in the BH masses as metallicity decreases.

The correlation between the metallicity and mass of WDs has important implications for detecting and characterising GWs from WDB as well, as it can affect whether they can be observed by GW detectors. For both BPASS and BSE the distributions are relatively smooth. However for BPASS we see a sudden shift at $\log(Z/Z_\odot)\approx~0.7$. We see a different trend in Figure \ref{fig:wd_Z}, the $\mathcal{M}_c$ of WDBs also correlates with their progenitor systems' metallicity. Specifically, higher metallicity systems tend to produce less massive WDs, leading to smaller cores and lower $\mathcal{M}_c$ in resulting binary systems. This trend is also observed, although more weakly, in the BSE WDB population.

\section{Synthetic LISA observations of WDBs and BHBs}
\label{section:3}

\subsection{Simulating LISA observations with \textsc{PhenomA} and \textsc{LEGWORK}}

With the population of WDBs and BHBs GW sources in a MW galaxy analogue we can estimate the signals from these objects, understand what LISA will eventually observe and which objects will be resolvable. To do this we make use of two publicly available packages for evaluating whether an object emitting GWs is detectable by LISA or not. These packages are \textsc{PhenomA} \citep{2007CQGra..24S.689A,2019CQGra..36j5011R} and the LISA Evolution and Gravitational Wave ORbit Kit \citep[LEGWORK,][]{LEGWORK_apjs}. 

We note that \textsc{PhenomA} has been updated to more generic and accurate models, such as the latest \textsc{PhenomB} model \citep{2014GReGr..46.1767H, 2015PhRvD..91b4043S}, which calculates spin-precession effects. However, \textsc{PhenomA}, is sufficient for calculating strains and estimating signal-to-noise ratios (SNRs) in our study. While \textsc{LEGWORK} is being increasingly widely used by the community. By using both analysis packages we are also able to evaluate if using different packages has any impact on evaluating how many sources in a binary population can be detected.

These packages work by first calculating the strain and amplitude spectral density (ASD), of a binary system based on its mass, orbital period and distance from LISA, see \citet{2007CQGra..24S.689A}, \citet{LEGWORK_apjs} or Appendix~\ref{waveform_calcs} for details. This is then compared to the LISA sensitivity curve which has two components. The first of these is the pure instrumental noise of the LISA detector, the second is the confusion noise from unresolved WDBs. The GW ADS from a binary source can then be compared to evaluate it's detectability, usually by assuming a minimum SNR. The details in our analysis are handled by \textsc{PhenomA} and \textsc{LEGWORK} and we detail the noise curve and SNR calculation in Appendix~\ref{noisecurve}. Both \textsc{PhenomA} and \textsc{LEGWORK} also output figures showing how the population of sources compare to the sensitivity curve as well as the number of systems that can be detected.

\subsection{The GW emission of the synthetic binary populations}
\label{3.2}

Before considering our mock observations of our synthetic WDB and BHB populations we should consider what actual observations exist that can help us evaluate the validity of our predictions. A significant amount of observational work has been undertaken to find LISA verification binaries \citep[e.g.][]{2018MNRAS.480..302K,2023arXiv230212719K,2023LRR....26....2A, 2023MNRAS.522.5358F}.

Verification binaries consist of a WD or neutron star primary and a compact subdwarf/WD/neutron star secondary that have been discovered by extant electromagnetic surveys. Most verification binaries have ultrashort orbital periods, less than an hour, producing GWs with frequency between $10^{-3}$ Hz and $10^{-2}$ Hz. Only a fraction of the BHB and WDB populations from both BPASS and BSE models fall in this frequency range. The strains of many of the verification binaries are accurately predicted and calculated as their distances have been measured by Gaia \citep{2018A&A...616A..11G}. The expected SNRs of these verification binaries are all $\geq 5$ for a four-year LISA mission time, with many SNR values above the widely used LISA SNR threshold of 7. Detailed information on current verification binaries can be found in \citet{2018MNRAS.480..302K}, \citet{2023arXiv230212719K}, \citet{2023LRR....26....2A}, \citet{2023MNRAS.522.5358F} and the ADS library\footnote{\url{https://ui.adsabs.harvard.edu/user/libraries/XOVGbNMAThua37oSEdV8LA}}.

\begin{figure*}
    \centering
    \includegraphics[width=2\columnwidth]{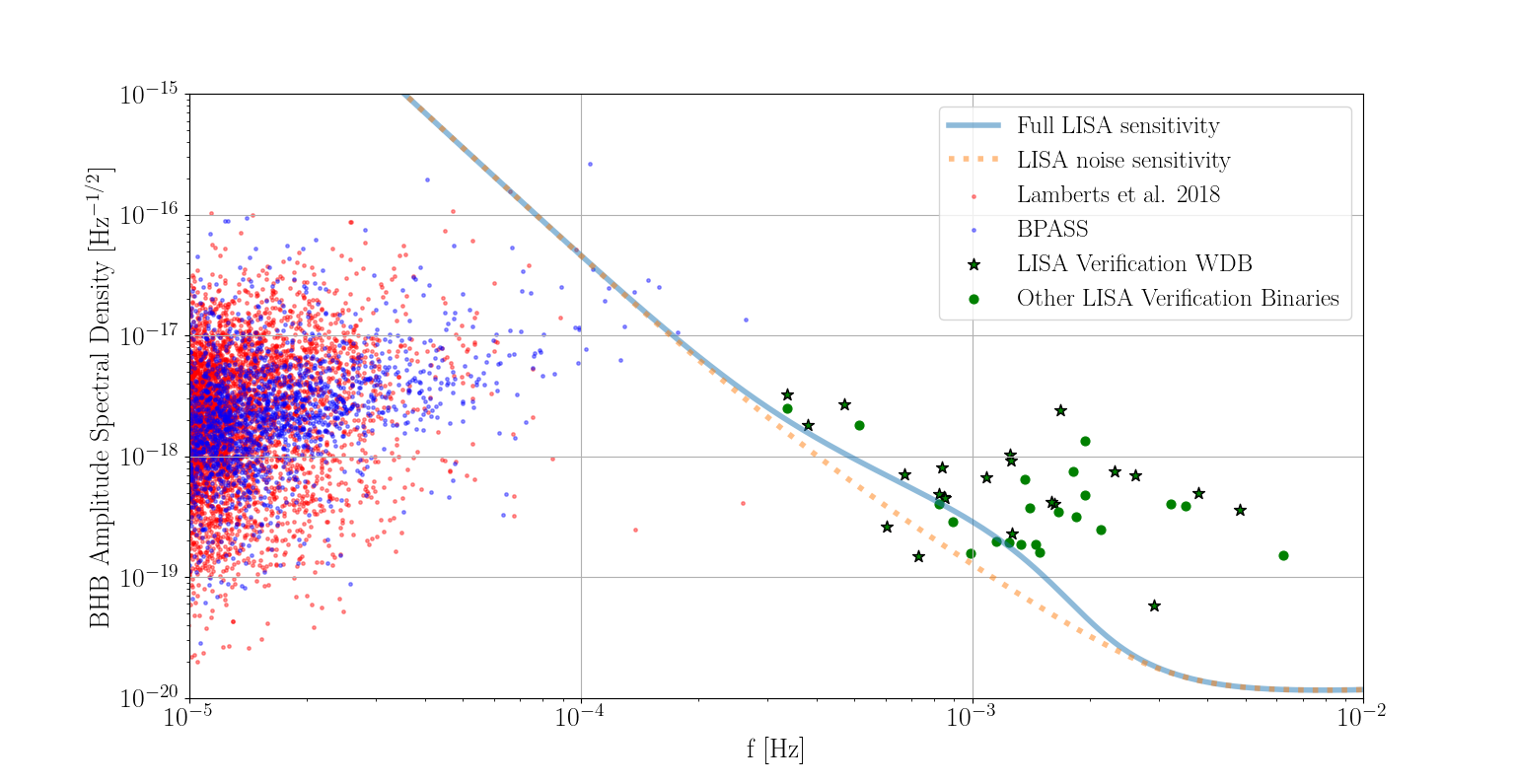} 
    \caption{The ASD for frequency range from $10^{-5}$ Hz to 0.1 Hz of the BHB populations as calculated by \textsc{PhenomaA}. The blue line is the approximated full sensitivity curve combining the instrument noise and the Galactic confusion noise for a four-year LISA mission lifetime. Blue points are individual BHBs from BPASS population and red points are BHBs from the BSE population. The model populations are observed from the Sun's location at (8, 0, 0). The black stars indicate the location of observed verification WDB and the green stars are verification binaries of other types.}
    \label{fig:bhbh_strain}
\end{figure*}

\begin{figure*}%
    \centering    
    \includegraphics[width=2\columnwidth]{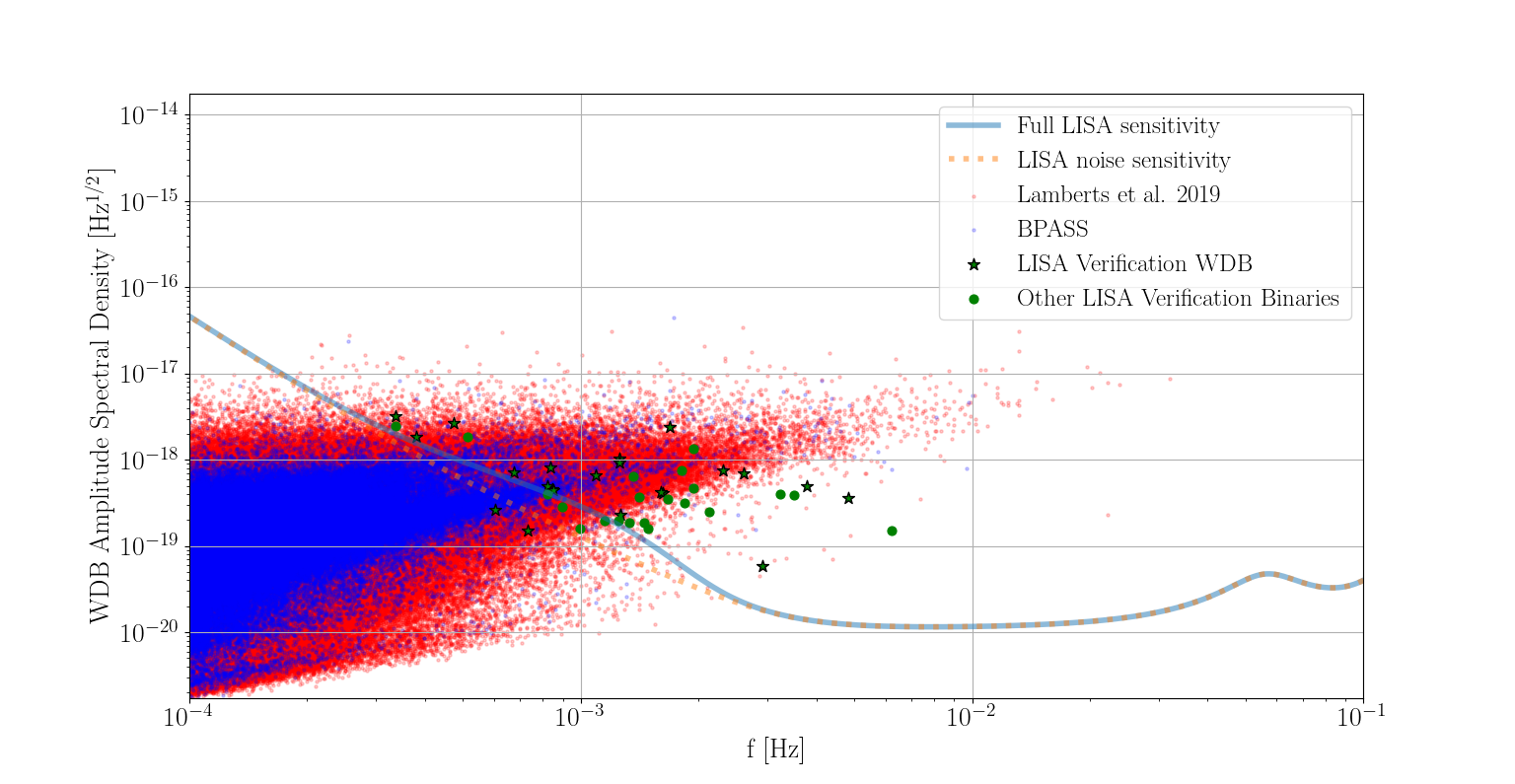}   
    \caption{ASD versus frequency for the WDB populations as calculated by \textsc{PhenomA}. The blue line is the approximated full sensitivity curve combining the instrument noise and the Galactic confusion noise for a four-year LISA mission lifetime. Blue points are individual WDBs from BPASS population and red points are WDBs from the BSE population. The model populations are observed from the Sun's location at (8, 0, 0). The black stars indicate the location of observed verification WDB and the green stars are verification binaries of other types.}%
    \label{fig:wdwd_phenoma}%
\end{figure*}

To visualise the ASD of individual Galactic binary sources, we plot our model BHB and WDB populations, the 43 known LISA verification binary sources' ASD (marked as green dots and stars) and the LISA ASD noise sensitivity curves with and without confusion noise in Figures~\ref{fig:bhbh_strain} and Figure~\ref{fig:wdwd_phenoma}. The ASD for the sources is calculated by using Eq.20--Eq.23 in \cite{2019CQGra..36j5011R} with the sensitivity curve ASD from equations B1 to B7. Many verification binaries on these Figures are well above the sensitivity curves, and thus LISA is expected to detect many of these individual verification binary sources \citep{2006CQGra..23S.809S, 2020ApJ...905...32B, 2021NatAs...5.1052P}. However, several are closer to the sensitivity curve and will be more challenging to resolve.

First, we notice most of the BHB population is below the sensitivity curve and will not be seen by LISA. Secondly, for both BSE and BPASS populations there are only a handful of BHBs above the sensitivity curve. BPASS models predict more BHBs with frequencies above 10$^{-4}$~Hz than BSE, which can also be seen in Figure \ref{fig:bhbh_fMc}. It is only systems above this frequency that can be detected by LISA, while BSE predicts more BHBs overall as BPASS reaches higher frequencies it predicts more detectable BHBs. We note that \textsc{LEGWORK} predicts almost exactly the same results as shown in Figure \ref{fig:legwork}. With the difference frequency distributions of the BSE and BPASS populations clearly evident.

Figure~\ref{fig:wdwd_phenoma} shows the ASD of the WDB populations for each of the two models against approximations of sensitivity curve ASD as calculated by \textsc{PhenomA}. We note that the \textsc{PhenomA} was incorrectly assuming systems with frequencies above 10$^{-2}$~Hz were evolving during the LISA observations and would merge. We adjusted this part of the code so that all WDBs were treated as stationary sources. We notice that, in contrast to the BHB populations, there are orders of magnitude more WDBs above the LISA sensitivity curve which can potentially be individually resolved. There is also a great population of sources below the LISA sensitivity curve, the source of the confusion noise. We notice that there are more high-frequency and high-ASD (> $\rm 10^{-18} Hz^{-1/2}$) systems in BSE model than the BPASS model, even though BPASS has more systems overall. The reverse of what is found for BHBs. Again, we have used \textsc{LEGWORK} to calculate the ASD as shown in Figure~\ref{fig:wdwd_strain}; where again we see that the results agree with \textsc{PhenomA}.

When we compare the synthetic WBD populations to the sample of verification binaries we note that both BSE and BPASS do have systems in the strain-frequency space of the verification binaries. However, the synthetic populations predict many systems with lower ASD, which are likely to be similar to the verification binaries in mass and period but at greater distances, where they will be more difficult to detect in electromagnetic surveys.

\subsection{The detectability of the synthetic binary populations}

To count the number of WDBs and BHBs that can be detected in our synthetic populations we count how many are above the LISA sensitivity curve. This is calculated by both \textsc{PhenomA} and \textsc{LEGWORK} using the method detailed in Appendix \ref{snr}. In this work we use an integrated observation time of four years with an assumed threshold SNR of 7 to be resolved. We do not impose to have unique systems per frequency bin and discuss the implications of this below.

\begin{figure*}
    \centering
    \includegraphics[width=2\columnwidth]{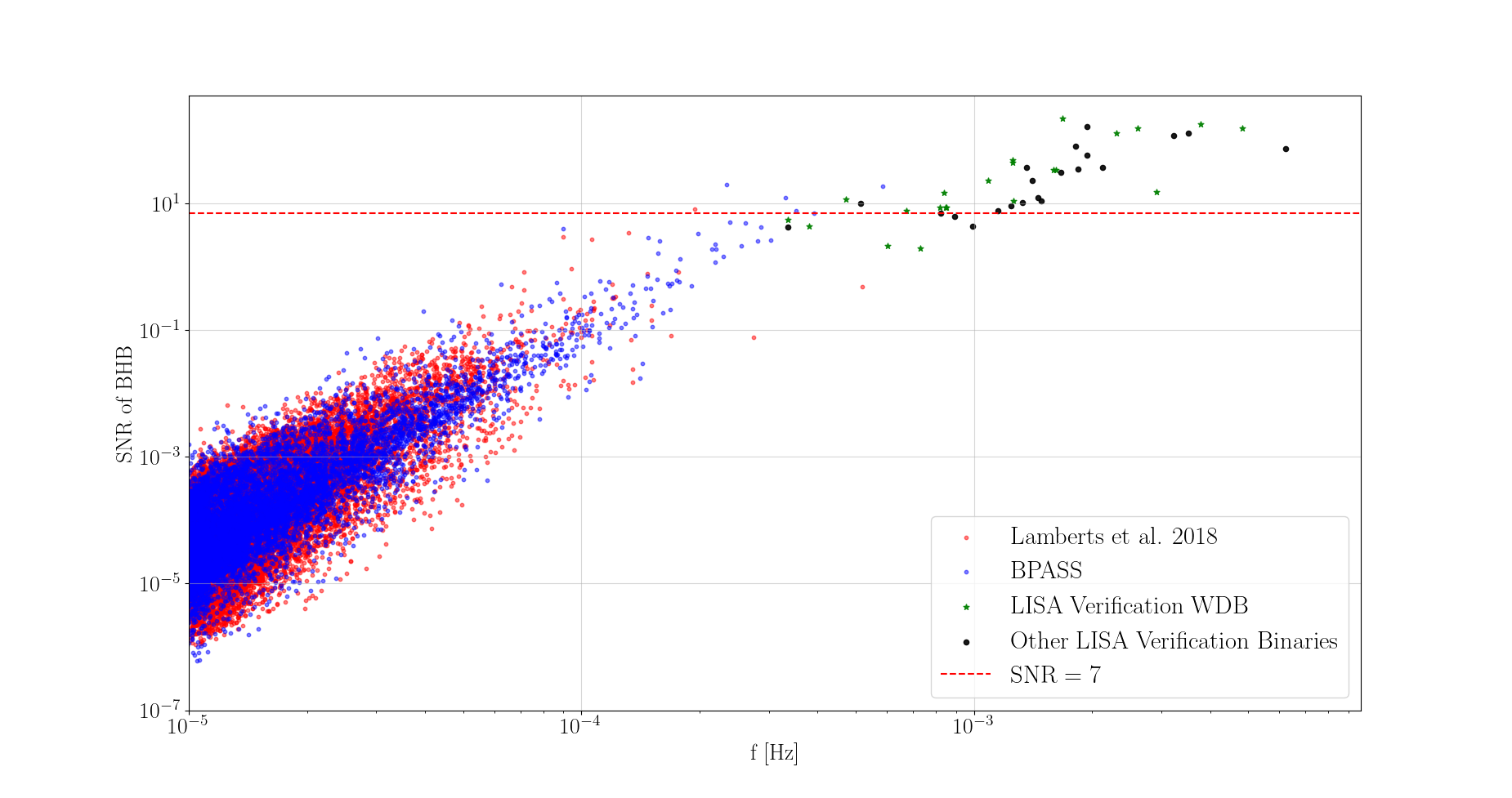}
    \caption{ The four-year mission SNR for the BHBs in the two model populations versus frequency. We also plot LISA verification binaries for comparison. The SNR threshold marked with red dashed line of SNR = 7.}
    \label{fig:bhbh_SNR}
\end{figure*}

\begin{figure*}%
    \centering
    \includegraphics[width=2\columnwidth]{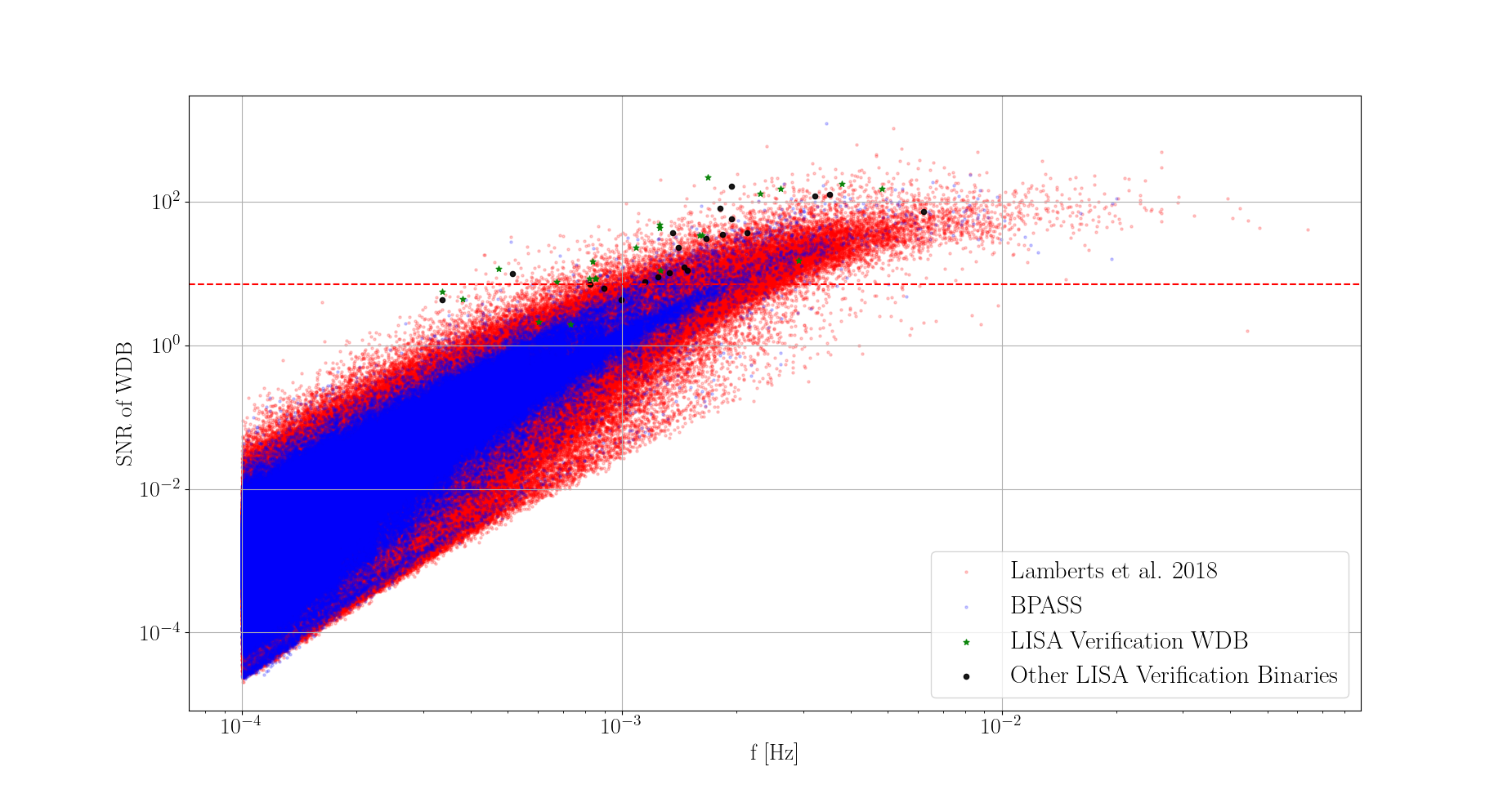}
    \caption{ The four-year mission SN for WDBs in the two model populations versus frequency. We also plot LISA verification binaries for comparison. Both populations predict systems similar to the verification binaries within the LISA frequency range. The SNR threshold of 7, which is represented by the horizontal red dashed line.}
    \label{fig:wdwd_snr1}
\end{figure*}

The SNR of sources vs their frequency are shown in Figures~\ref{fig:bhbh_SNR} and \ref{fig:wdwd_snr1}. With the number of systems above the SNR of 7 shown in Table \ref{tab:resolvable}. We see that both \textsc{PhenomA} and \textsc{LEGWORK} predict numbers of detected systems that agree closely. In this table we have also calculated the detected population at four different locations around the model galaxy to determine whether the position of LISA in the galaxy affects the population detected. 

Figure~\ref{fig:bhbh_SNR} displays the individual SNRs of BHBs in our two populations. BPASS has a greater population of high frequency BHBs. The result of this is that four sources are predicted to be detected from the BPASS model, compared to only one from the BSE model. We note the work of \citet{2020MNRAS.494L..75S} built on the work of \citet{2018MNRAS.480.2704L}, using the same BSE model but made 300 synthetic realisations of the BHB populations and predicted 4.2 detectable BHBs with the same SNR threshold of 7 and observation time. This suggests that while in the BSE population we employ our number of detected systems is 0.5 this might just be because of the distributions of BHBs in this model population rather than an inherent difference between BSE and BPASS. Also \citet{2020MNRAS.494L..75S} found that having one detected BHB was within the 90 per cent confidence interval. A large number of BHB predictions are compiled by \citet{2022ApJ...937..118W}, with predicted BHB detections from zero to 154, although most studies predict a few BHB detections. Thus at least in the high mass regime BPASS is consistent with other work.

We also plot predictions of individual SNRs of Galactic WDBs in two populations in Figure~\ref{fig:wdwd_snr1}. The two WDB populations overlap each other most at low frequencies. The BSE model predicts 13,314 WDB sources above the threshold of 7, and BPASS predicts 673 WDBs with a SNR greater than 7. The BPASS WDB population has more systems at lower frequencies. However, BPASS has fewer individually observable WDB systems than that of the BSE WDB population. This significant difference between the BPASS and BSE WDB are discussed below. The BSE results are more similar to other predictions \citep[e.g.][]{2001A&A...375..890N} where it is expected that 10,000s of WDBs will be resolved by LISA.

The effect of the choice of the Sun's position on the number of individually resolved systems is also tested, and results are shown in Table~\ref{tab:resolvable}. The predictions from different choices for the Sun's position in the galaxy do not vary significantly but they allow us to estimate the uncertainty in the detected number of sources due to where LISA is positioned within the Galaxy.

In Table~\ref{tab:resolvable} we have also included the number of sources that can be detected if we remove the confusion noise. Give than with BPASS we predict a factor of 20 fewer WDBs it is highly likely that the confusion noise from unresolved binaries will be correspondingly lower. Without the confusion noise the number of observable WDBs increases to 1048. This indicates that for a smaller population of WDBs, more individual sources can be detected compared to the populations typically predicted by other studies.

\begin{table}
  \centering
  \caption{ Number of Galactic binary systems with SNR > 7
  in the MW-like galaxy predicted by different models, for different locations for the Sun in the galaxy for \textsc{PhenomA}  (top half) and \textsc{LEGWORK} (bottom half).}
  \begin{tabular}{lcccccccc}
  \hline
  \hline
     & \multicolumn{4}{c}{\textsc{PhenomA}} \\
     \hline
   & \multicolumn{2}{c}{BHB} & \multicolumn{2}{c}{WDB} \\
    $\rm Location_\odot$ & BSE  & BPASS &  BSE & BPASS \\
\hline (8, 0, 0) &-- & 4 & 13,296 & 682 \\
       (-8, 0, 0)& 1  & 5 & 13,375 & 676  \\
       (0, 8, 0)& 1  & 4 & 13,187 & 665 \\
       (0, -8, 0)& -- & 5 & 13,401 & 681  \\
\hline Mean & 0.5$\pm$0.6 & 4.5$\pm$0.6 & 13,315$\pm$96 & 676$\pm$8 \\
\hline
\hline
     & \multicolumn{4}{c}{\textsc{LEGWORK}} \\
\hline
       (8, 0, 0) & -- & 4 & 13,882 & 678 \\
       (-8, 0, 0)&  1 & 5 & 13,893 & 669  \\
       (0, 8, 0)& 1 & 4 & 13,505 & 660 \\
       (0, -8, 0)& -- & 5 & 13,959 & 674  \\
\hline Mean & 0.5$\pm$0.6 & 4.5$\pm$0.6 & 13,810$\pm$206 & 670$\pm$8 \\
\hline
Combined mean & 0.5$\pm$0.5 & 4.5$\pm$0.5 & 13,562$\pm$304 & 673$\pm$8\\
\hline
\hline
     & \multicolumn{4}{c}{\textsc{PhenomA} - no confusion noise} \\
     \hline
   & \multicolumn{2}{c}{BHB} & \multicolumn{2}{c}{WDB} \\
    $\rm Location_\odot$ & BSE  & BPASS &  BSE & BPASS \\
\hline (8, 0, 0) & --& 4 & 19,156 & 1,036 \\
       (-8, 0, 0)& 1  & 5 & 19,154 & 1,062  \\
       (0, 8, 0)& 1  & 4 & 18,867 & 1,048 \\
       (0, -8, 0)& -- & 5 & 19,281 & 1,045  \\
\hline Mean & 0.5$\pm$0.6 & 4.5$\pm$0.6 & 19,115$\pm$175 & 1,048$\pm$11 \\
\hline
  \end{tabular}
  \label{tab:resolvable}
\end{table}

When combined, the results confirm previous findings that WDBs and BHBs will be detected by LISA but the true number is still uncertain. Unfortunately, it appears that significant uncertainties remain in the modelling of the binary evolution itself. Even when choosing different locations within the Galaxy the number of detected signals does not significantly vary. Although the model of the galaxy itself will have a significant impact on the predictions. Thus, when preparing for LISA and its data analysis, flexibility must be included because until it flies we cannot be certain about the signals it will detect beyond the known verification binaries.

If we compare to the tabulated results of \citet{2022ApJS..260...52W}, we find that our predicted BHB population of a few is typical of other predictions (see Figure 12 in \citet{2022ApJ...937..118W}), although they suggest that up to 154 BHBs (for certain model parameters) could be detected! This is because \citet{2022ApJ...937..118W} widely varied the input physics in the COSMIC population synthesis code (an updated, user friendly version of BSE\footnote{\url{https://cosmic-popsynth.github.io/COSMIC/}}). It is true that this gives an insight into the uncertainty of the binary population synthesis, but some of those parameters would be ruled out by observations. BPASS has been repeatedly validated against numerous different observations \citep[i.e.][]{2016MNRAS.457.4296W,2017PASA...34...58E, 2018MNRAS.479...75S, 2019MNRAS.482..384X, 2019MNRAS.482..870E, 2021MNRAS.507..621B,2021ApJ...922..177M}. While the physics in BPASS can still be refined the repeated agreement with observed stars gives confidence in results. However, in the low mass regime of WD evolution, this is one area where further testing must occur, and our results for the WDB population are less certain.

We further remark that when considering the detectability of a system in the LISA frequency band, it is important to consider the SNR and the presence of other binaries within the same frequency bin; especially, the number of systems is high in the lower LISA frequency bins \citep[see][]{2020ApJ...898...71B}. In the LISA frequency band, each binary system emits GWs at a specific frequency, assuming the signals are stationary during the 4-year LISA mission time), but the LISA instrument has a finite frequency resolution (width of each frequency bin). Even if a binary system has a sufficiently high SNR > 7 to be detected above the noise, it is still crucial to consider the presence of other binaries within the same frequency bin. A full detection pipeline is needed. Suppose another binary system has a similar or identical emission frequency within that bin. In that case, it may be challenging to distinguish the individual signals from each other and from the background noise. In such a scenario, the multiple overlapping signals could interfere with each other, making it difficult to extract the individual signals. This effect is particularly relevant when numerous binaries emit GWs within a specific frequency bin, which is possible given the Galaxy's large expected population of binary systems. This effect was discussed in more detail in \citet{2023MNRAS.524.2836V}. As a result, the number of detectable binaries may be reduced because the overlapping signals hinder the ability to separate them from the noise and the other binaries within the same bin. We need to consider the SNR and the potential interference caused by multiple binaries within the same frequency bin when predicting the detectability of GWs in the LISA frequency band. This consideration could help provide a more realistic estimation of the number of observable binaries in our research. The size of this can be estimated by comparing our number of detected sources in our study, of 13,500 WDBs to the 12,000 detected by \citet{2019MNRAS.490.5888L} who included this effect. That suggesting that approximately 11~per~cent fewer systems may be detected.

\section{Detection of synthetic binary populations and Creation of the Mock-LISA Signal}
\label{4}
\label{4.2}

Now that we have our population of WDBs and BHBs, our next step is to predict the GW signals that LISA will detect from all the sources at once. The aim is to create the ASD of the Galactic foreground signal from WDBs and BHBs that will be observed over a year. To achieve this we use the method of \citet{2021MNRAS.508..803B}. The full details are described in Appendix~\ref{waveform_calcs}.

We demonstrate in previous sections the detectability by LISA of different binary populations in model galaxies. In this section, we present the results of our modulated signal predictions in time domain. This is to show that we can produce such signals, which will lead to our next data analysis paper. In creating these modulated signals we include all WDBs and BHBs with a frequency of GWs above 10$^{-5}$~Hz and we use the method outlined in Appendix~\ref{waveform_calcs}.

The motion of LISA around the Sun causes a periodic variation in the GW signal detected by LISA, because of the Doppler effect associated with this motion. This variation has a characteristic pattern called a "sine-Gaussian" modulation, from which one can extract information about the binary system, such as sky position, distance, orientation, and orbital parameters, and study the MW galaxy. The potential to extract information about binary systems makes studying the modulation important. 

\begin{figure*}
    \centering
    \includegraphics[width=2\columnwidth]{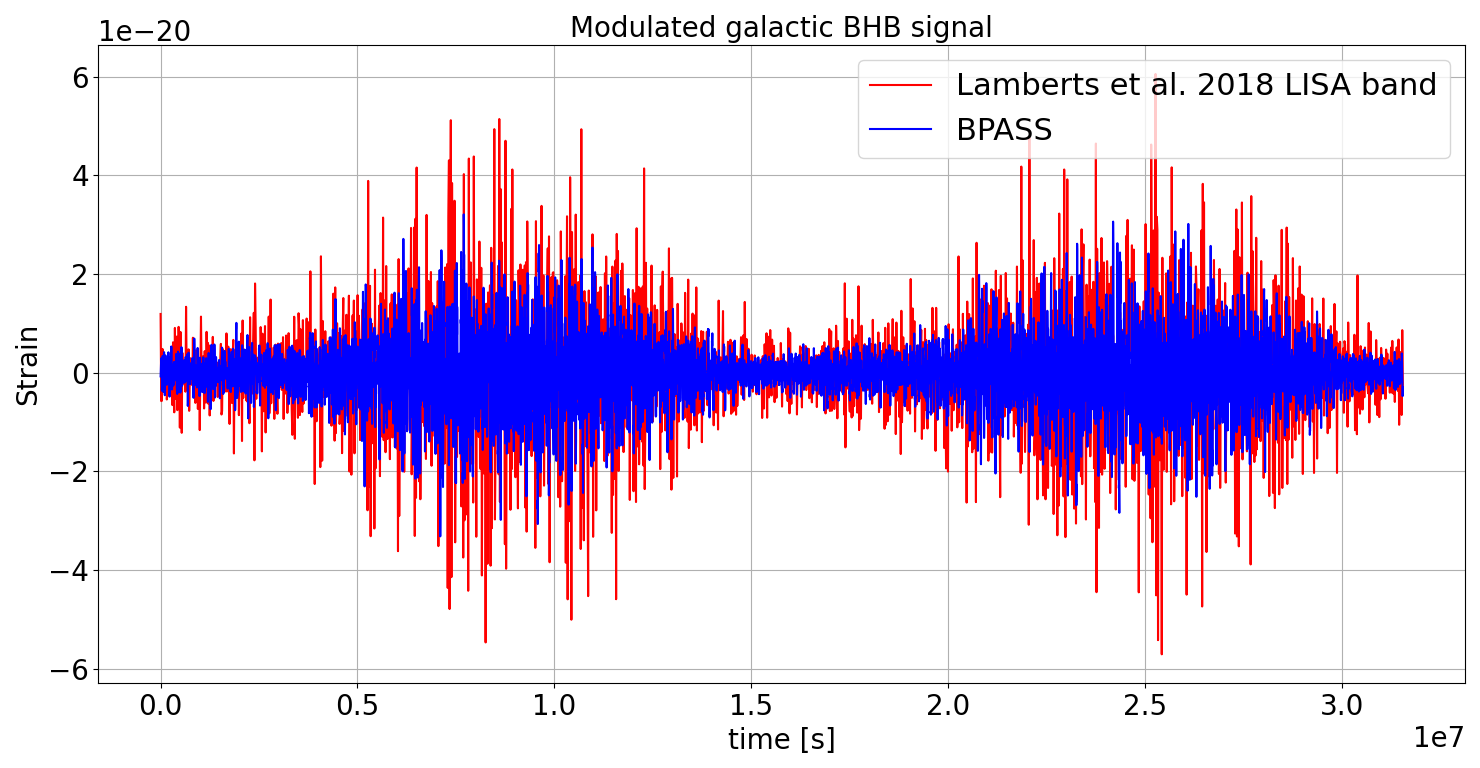}
    \caption{Modulated GW strain signals of Galactic BHB populations are shown over the duration of a one-year LISA mission. The blue line represents the sum of 9,298 BPASS Galactic BHBs, and the red line represents the sum of 25,335 BSE Galactic BHBs. }
    \label{fig:modulated_bhbh}
\end{figure*}

\begin{figure*}
    \centering
    \includegraphics[width=2\columnwidth]{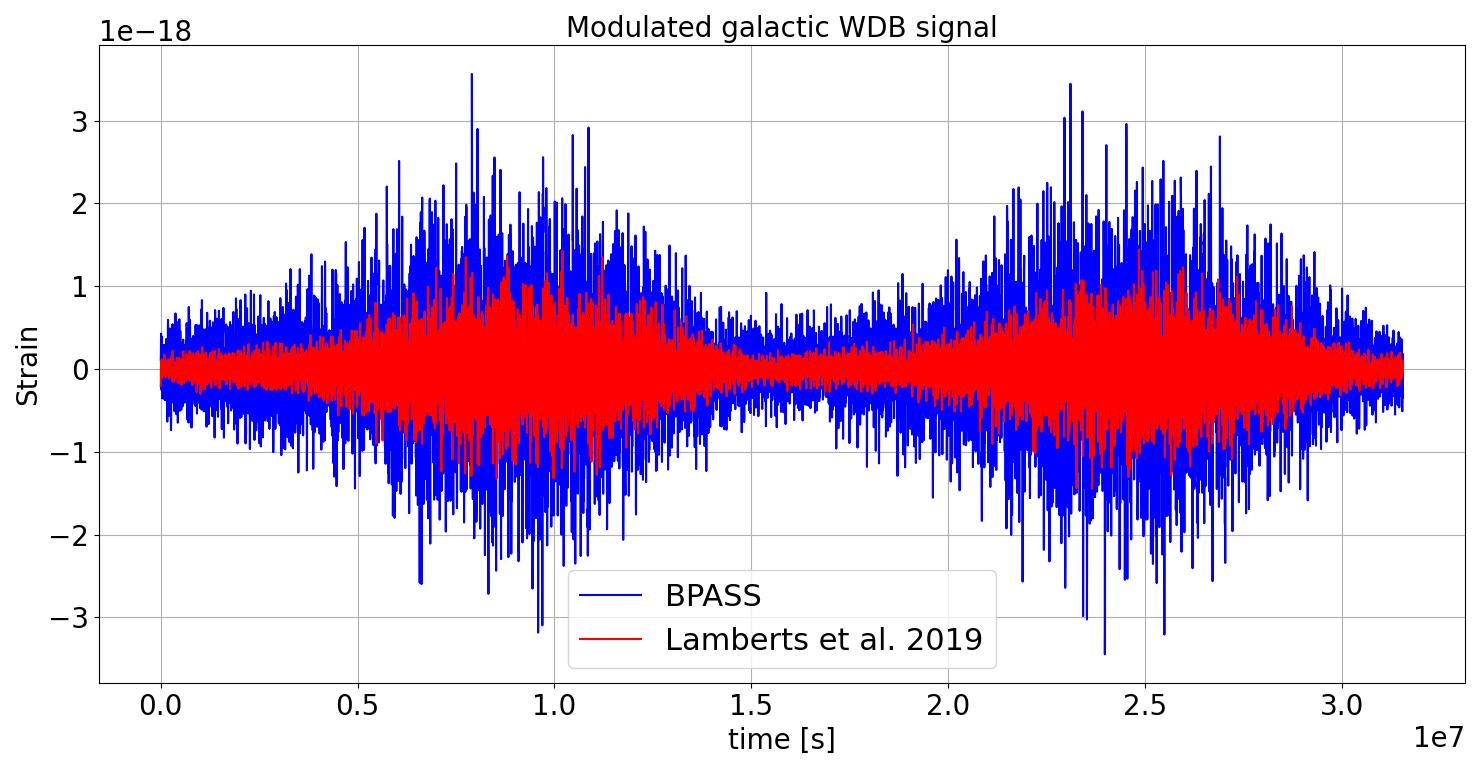}
    \caption{Modulated GW strain signals of Galactic WDB populations are shown over the duration o a one-year LISA mission. The blue line represents the sum of 58,522,007 BPASS Galactic WDBs, and the red line represents the sum of 35,832,029 BSE Galactic WDBs. }
    \label{fig:modulated_wdwd}
\end{figure*}

From our simulated binary populations, we produce and predict a modulated GW signal for each binary in the Galaxy according to Equation~\ref{eqn:modulated}. The time variable changes linearly from 0~s to $3.16\times 10^{7}$~s or 1 year, with a time step of 771~s. Actual LISA data will be recorded with a much smaller sampling frequency. With a 771~s time step we require $2^{12}$ data points, which are sufficient to produce the modulation without requiring too much computational time. To produce the linear combination modulation we sum up all the individual signals in the galaxy to produce the final modulated signals for a particular binary population (using Equation~\ref{combinedmodulatedsignal}). We show the mock time domain signals in Figures~\ref{fig:modulated_bhbh} and \ref{fig:modulated_wdwd}. Comparing the BHB and WDB populations, the BHB strain signal is about one order of magnitude stronger than that of the WDB population, and this difference is to be expected as the GW strain is proportional to the mass of the binary system.


Both Figures~\ref{fig:modulated_bhbh} and~\ref{fig:modulated_wdwd} show modulated signals, as the measured wave amplitude changes over the LISA orbit. Physically, modulation occurs when LISA faces the Galactic centre where the population is concentrated. BPASS and BSE predict that the modulated signal from the WDB is about two orders of magnitude stronger than that of the BHB. Figure~\ref{fig:modulated_bhbh} shows the that the maximum strain of the modulated BHB population strain reaches to just below $2\times 10^{-20}$ for the BPASS model, compared with that of BSE, which is double this value.  In comparison the reverse is true for WDBs. The strain of the modulated WDB population signal for the BPASS model is about twice that predicted by BSE, shown in Figure~\ref{fig:modulated_wdwd}. For both modulated signals the amplitude is greatest for the population with the greatest number of sources, rather than being dominated by the highest strain systems. 

The differences between BPASS and BSE are expected as BPASS predicts seven times more Galactic WDB than BSE. While BSE predicts 2.5 times the number of BHBs than BPASS. Because the total strain signal is the sum of all the signals the more numerous the sources, the higher the combined strain. This is also demonstrated in Figure~\ref{fig:modulated_wdwd3} where we show a total strain including WDBs with frequencies below 10$^{-5}$~Hz which increase the strain further.

These results show that the synthetic signals are sensitively linked to the binary evolution model used. Furthermore, their features are linked in a non-linear way to how many sources are detectable. It is important to note, because analysis pipelines that take this data and analyse it, must have the flexibility to allow for a wide range of possible signals.

\section{Discussion}
\label{section:5}
In this article we have combined the FIRE high-resolution cosmological simulation and BPASS binary population predictions for GW sources. This has allowed us to predict the number and parameter distributions of unmerged Galactic BHBs and WDBs in a simulated MW-like galaxy. From this, we have simulated the expected modulated LISA signals and identified the expected population of sources that can be resolved with LISA. Our predictions will aid in developing data analysis methods, we show how this will be done in a future paper.

The GW population prediction is sensitive to the choice of binary population synthesis model. We compared our results with the previous work of \citet{2018MNRAS.480.2704L,2019MNRAS.490.5888L} who used the same FIRE galaxy simulation but the BSE binary population synthesis code. This allows us for the first time to see how sensitive such predictions are when only the population synthesis code used is varied. We find significantly different populations of WDBs and BHBs, showing that choice of stellar models introduces a significant uncertainty on the predicted compact remnant binary population.

The differences are apparent in all aspects of the results; not just in the total number of sources but in the distribution of masses, frequency, metallicity and the number of systems that can be resolved by LISA. It is key to validate these binary population synthesis codes by comparing to other observations. Multiple observational constraints should be used to counter degeneracies between the input physics and how it affects predictions. We note that the BPASS models have been widely tested against many varied previous observational constraints. However, there were historically relatively few comparisons to lower mass stars relevant for WDBs, though we have begun to explore this regime \citep[e.g.][]{2022MNRAS.512.5329B,2022MNRAS.514.1315B}.

When we compare the BSE and BPASS, BPASS predicts more detectable BHBs than BSE despite predicting fewer BHBs overall. This is because the mass and frequency distribution of BPASS BHBs extends to higher chirp masses and frequencies. Although we note that \citet{2020MNRAS.494L..75S} using more realizations of BSE did find on average that 4.2 BHBs would be detected. We also note that BPASS also reproduces the high mass end of the GW transient chirp mass distribution \citep[e.g.][]{2023MNRAS.520.5724B}. Which suggest BPASS prediction for LISA BHBs are closer to what might be observed. However, we note there are of course other pathways such as dynamical interactions in clusters to form GW transients.

For the WDB population, BPASS predicts more WDBs in total but fewer observable systems than BSE. BPASS has not been validated for such low-mass stars as it has been for high-mass stars. So for WDBs it is difficult to be certain as to which population is more realistic, at least until more complete catalogues for known WDBs from electromagnetic observations are available. However, we note that BPASS can reproduce the observed type Ia rate and it's evolution with redshift \citep[detailed in][]{2019MNRAS.482..870E,2022MNRAS.514.1315B}. However, the majority of Type Ia SNe would then be via the single-degenerate rather than double-degenerate channel. 

Binary evolution is uncertain and the different populations found for the WDBs here show that different input physics can lead to order of magnitude differences in the predicted number of LISA sources. Further work to understand these differences will require the comparison of a large number of the extant binary population synthesis codes as well as varied observations to constrain them until LISA launches and takes its first observations.

This study is one of the first to directly compare the difference in outputs between a binary stellar evolution models based on BSE rapid stellar evolution \citep[][]{2002MNRAS.329..897H} and that from the BPASS detailed stellar evolution models \citep[][]{2018MNRAS.479...75S}. It is an interesting comparison because the equations that represent the stellar evolution for BSE \citep[][]{2000MNRAS.315..543H} were based on detailed stellar evolution models calculated by the Cambridge STARS code originally described in \citet{1971MNRAS.151..351E} with the last paper in common between the two being \citet{1995MNRAS.274..964P}.

Given this fact one might expect that the results from the two codes would be identical for stellar populations. But they are clearly not. The reason for the differences are complex. While for single stars the two codes must create similar results due to the same stellar physics being used. Even with two differences between the BPASS version of the Cambridge STARS code and that of  \citet{1995MNRAS.274..964P}. BPASS uses newer opacity tables \citep[][]{2004MNRAS.348..201E} and an updated stellar wind mass-loss prescription \citep[][]{2017PASA...34...58E}. The differences from this however only impact on the latest stages of stellar evolution and the BHBs from massive stars. Thus differences are thus likely to arise from the different ways binary interactions are dealt with.

The binary physics model of \citet{2002MNRAS.329..897H} was used as a guide for the physics of BPASS but several modifications had to be made because of the nature of how detailed stellar evolution models work. For example, it is not possible to just remove the hydrogen envelope in a detailed model, it must be done over a number of timesteps. Thus, the CEE prescription is inherently different. The implementation of the CEE within BPASS also assumes angular momentum conservation rather than the typical conservation of energy that is employed in BSE \citep[see the supplementary information of][for further discussion on the different CEE prescription]{2023NatAs...7..444S}. Thus one possible explanation for fewer tight WDBs in BPASS is that the CEE model is not producing enough tight binaries. We have made some recent exploratory models that indicate tighter WDBs can be made if a BSE-like CEE prescription is implemented within BPASS so this is the most likely cause of the different between BPASS and BSE WDB populations. However, creating a full population to test this would require to recalculate a new series of BPASS stellar evolution models with this different CEE scheme. This would be highly computationally expensive but is clearly something we must further investigate. 

Another difference that is more another important source of the differences between the BSE and BPASS results is the relative rates of SMTs. In BSE mass transfer stability is determined by analytic expressions comparing the evolutionary state of the stars and their mass ratio \citep[][]{2002MNRAS.329..897H}. In BPASS when the donor star fills it's Roche lobe, mass is lost from the donor and the donor star then either reaches an equilibrium so that the stellar radius is stable, or the stellar radii continues to grow so that CEE occurs. Thus in BPASS how the star responds to mass loss determines what happens to the binary. We find, especially at lower masses, MT is much more stable than would be assumed in BSE, leading to a much lower occurrence of CEE. Thus even though the CEE models are different it is the fact that SMT is more prevalent that could be the reason for much of the difference. The clearest indication this is the case is shown in Figure~\ref{fig:wdwd_m12}. Here the fact so many of the secondary WDs are more massive than the primary WDs is due to efficient MT during the first occurrence of MT within the binary. Further implications of the prevalence of SMT in BPASS are discussed in \citet{2023MNRAS.520.5724B}.

A final difference between BSE and BPASS comes from a more subtle result of the use of detailed stellar evolution models. MT can occur at almost any phase of a star's evolution. How the core mass and core convective zones react to changes in a star's mass can be accurately followed in a detailed stellar evolution code. MT also ends when a star shrinks within its Roche lobe rather than assuming a star must lose it's whole hydrogen envelope. These differences in how the stellar models evolve could be the reason behind the different structures in the chirp mass--GW frequency distributions of Figure~\ref{fig:wdwd_fMc}. Especially at the lower chirp masses where BSE does not allow some WDBs to exist while they are predicted in the BPASS results.

All of these factors lead to the different distribution of WDBs from BPASS that we find compared to that from \citet{2018MNRAS.480.2704L,2019MNRAS.490.5888L}. In particular, we find that there are fewer CEE events and mergers from stars with initial masses below approximately 5~M$_{\odot}$. These low mass stars have much slower evolutionary timescales and with our method of determining MT stability this leads to mass transfer being much more stable as the stars do not continue to grow in radius when they fill their Roche lobes. This increased MT stability means they avoid CEE, leading to fewer very close orbits and fewer binaries merging.

Further analysis of the source of the difference between the BPASS and BSE binary evolution algorithms is beyond the scope of this work. However, we can draw a firm conclusion that the treatment of physical processes within binary population synthesis codes does have a significant impact on the predicted WDB and BBH populations. This indicates that the future observations by LISA will play a important role in constraining our understanding of those physical processes.

\section{Conclusion}
\label{section:6}

We summarise our most important findings in this paper as follows:

\begin{enumerate}

\item Our simulations predict a population of BHBs in the MW. The BPASS population is smaller than that of \citet{2018MNRAS.480.2704L}. However, the number of detectable BHBs, with a SNR greater than 7, is greater. We expect to detect 4.5$\pm$0.5 from BPASS predictions compared to 0.5$\pm$0.5 from BSE. Our prediction is similar to \citet{2020MNRAS.494L..75S}, whom predicted 4.2 detectable BHB systems from using a large number of BSE model WDB populations. Our values also agree with the simulations collated by \citet{2022ApJ...937..118W}. With the range of values found by other works ranging from 0 to 154.

\item Our results confirm that there will be a significant overall population of WDBs in the MW galaxy, greater than the previous study \citep[][]{2019MNRAS.490.5888L} found. The predicted distribution of masses and separations of these WDBs is consistent with observations of LISA verification binaries and the extrapolated  double-degenerate WDB population. However, due to the verification binaries being an inhomogenous sample, it is difficult to compare our predicted number of such binaries beyond 10 kpc. 

\item We predict 673$\pm$8 LISA resolvable Galactic WDBs from BPASS compared with 13562$\pm$304 from BSE that have a SNR greater than 7 and would be able to be resolved by LISA (see Table~\ref{tab:resolvable}). 

\item With the smaller number of WDBs from BPASS it is likely that the confusion noise from undetected WDBs will be substantially weaker. If we assume this means we can ignore the confusion noise BPASS then predicts up to $1048\pm11$ WDBs could be detected.

\item We find our prediction for WDB systems with SNR greater than 7 is only weakly dependent on the tool used to determine detectability. With predicted numbers varying a few per cent between \textsc{PhenomA} and \textsc{LEGWORK}. We see no difference in our BHB numbers but there are very few objects detected in this population.

\item We find the choice of binary population synthesis codes introduces significant differences in the predicted compact remnant binary population. This is starkly revealed in Figures~\ref{fig:bhbh_strain} and \ref{fig:wdwd_phenoma}. This is likely due to differences in how the binary interactions are dealt with in BSE and BPASS, as well as due to the use of rapid stellar evolution modelling in BSE compared to the detailed modelling of the stellar evolution in BPASS. Further comparison to low-mass binary stars with WDs will be key to revealing which synthetic binary population is closer to what LISA will be expected to observe. We also need to evaluate the different pathways that lead to WDBs and see what impact extra physics such as magnetic-wind braking by white dwarfs may have on our predicted populations \citep[e.g.][]{2019MNRAS.483.5595V}. We do find that the total number of predicted WDBs by BPASS is similar to what is expected based on the density of WDBs in the Solar neighbourhood.

\end{enumerate}

Finally, we note that the simulation-based method we have used provides a powerful tool for predicting the population of BHBs and WDBs in a MW mass-like galaxy. Our predicted populations are consistent with other observational constraints, but the low-mass models need more validation. This is difficult because binary population synthesis codes have now been extant for decades, and providing strong constraints of the evolution of binary star populations is still somewhat uncertain. In this work, we have demonstrated that LISA observation will provide an important constraint on binary populations. As detailed in the LISA Mission L3 proposal \citep[see details in recent report by][]{2024arXiv240207571C}, the data processing for LISA involves several distinct levels of analysis to extract meaningful information from the raw data. The insights and mock data we have created here may have use as future LISA data flows are designed (at levels 2 and 3).

\section*{Acknowledgements}

We thank Max Briel for useful discussions on MT in BPASS models. We gratefully acknowledge support by the Marsden Fund Council from New Zealand Government funding, managed by Royal Society Te Apārangi, the University of Auckland and the University of Warwick for their continuing support. BPASS is enabled by the resources of the NeSI Pan Cluster. New Zealand’s national facilities are provided by the NZ eScience Infrastructure and funded jointly by NeSI’s collaborator institutions and through the Ministry of Business, Innovation \& Employment’s Research Infrastructure programme. URL: \url{https://www.nesi.org.nz}. AL acknowledges support by the ANR COSMERGE project, grant ANR-20-CE31-001 and the ``Programme National des Hautes Energies'' (PNHE) of CNRS/INSU co-funded by CEA and CNES. GB acknowledgres support from the Centre National D'\'{E}tudes Spatiales (CNES).

\section*{Data Availability}
The BPASS Galactic binary populations used in this article will be shared on reasonable request to the corresponding author. For more a general BPASS output regarding binary evolution please visit \url{https://bpass.auckland.ac.nz} or \url{http://warwick.ac.uk/bpass}.
The \textsc{PhenomA}  model used in this study can be download from the GitHub reference in \cite{2019CQGra..36j5011R}.\\
The LEGWORK model used in this study can be download from GitHub, and the detail of the software is outlined in \cite{LEGWORK_joss}.

\bibliographystyle{mnras}
\bibliography{mnras_main}

\newpage
\appendix

\section{GW Waveform Calculation}
\label{waveform_calcs}
For the benefit of readers interested in exploring the data and results of this study in greater depth, this appendix presents mathematical formulations, alternative versions of several key plots and supplementary plots included in the main body of the paper. These alternate plots offer additional insights and perspectives that enhance the reader's understanding of the research while providing a transparent and comprehensive view of the study's findings.

The first step in calculating the GW signal from a compact binary is to predict the waveform from these objects. In a MW-like galaxy, the distance between a binary system of a Cartesian coordinate $(X, Y, Z)$ and LISA $(X_\odot,Y_\odot,Z_\odot)=(8,0,0)$ in kpc is
\begin{equation}
    R = \sqrt{(X-X_\odot)^2+(Y-Y_\odot)^2+(Z-Z_\odot)^2}.
\end{equation}

The strain of the GW signal is proportional to the amplitude of the signal; and the GW amplitude depends on the mass of the binary and the distance between the source and the detector. Similar to electromagnetic radiation, GWs also come with two polarisations, namely plus (orthogonal) and cross (diagonal). The amplitude of the plus and cross-polarized GW emitted by a binary is \citep{2021MNRAS.508..803B}
\begin{equation}
    A_{+}(M_{1},M_{2},R,f,\iota)=\frac{2G^{2}M_{1}M_{2}}{c^{4}R}\left(\frac{(\pi f)^{2}}{G(M_{1}+M_{2})}\right)^{1/3}(1+\cos^{2}(\iota))
\end{equation}
\begin{equation}
\label{eq:A_x}
    A_{\times}(M_{1},M_{2},R,f,\iota)=-\frac{2G^{2}M_{1}M_{2}}{c^{4}R}\left(\frac{(\pi f)^{2}}{G(M_{1}+M_{2})}\right)^{1/3}2\cos(\iota).
\end{equation}
where  $M_{1}$ and $M_{2}$ are primary and secondary masses, in solar mass $M_\odot$, of the binary. $G$ is the Gravitational constant. $f$ is the GW frequency, which is twice the binary's orbital frequency for circular orbits. 
The inclination $\iota$ is the angle between the orbital plane of the binary and an observer. A $0^{\circ}$ inclination means the GW is travelling face-on towards LISA and has maximum strain signal. Assuming maximum inclination then  a factor of 4/5  needs to be included in the Equation~\ref{eq:A_x} if one wants to consider the average strain over all inclination angles \citep{2019CQGra..36j5011R}. Studies such as \cite{2023MNRAS.522.5358F} have shown from their models that different inclinations can affect amplitudes estimation. Another method is to draw $\cos(\iota)$ from a uniform distribution as there is no way to measure individual binary star inclination \citep{2021MNRAS.508..803B}. In our models, we assume $cos(\iota)$ is randomly drawn from a uniform distribution between -1 and 1.

The time-dependent GW strain $h(t)$ is given by the linear combination of two polarisations of the waveform,
\begin{equation}
    h(t)=h(t)_{+}{\pmb e}_{+}+h(t)_{\times}{\pmb e}_{\times},
\end{equation}
where ${\pmb e}_{+}$ and ${\pmb e}_{\times}$ are the two polarisation tensors, which are given by
\begin{equation}
  {\pmb e}_{+}={\pmb E}
  \begin{pmatrix}
    1 & 0 & 0 \\
    0 & -1 & 0 \\
    0 & 0 & 0
  \end{pmatrix}{\pmb E}^{T}\quad
  {\pmb e}_{\times}=E
  \begin{pmatrix}
    0 & 1 & 0 \\
    1 & 0 & 0 \\
    0 & 0 & 0
  \end{pmatrix}{\pmb E}^{T}.
\end{equation}

The polarisation coordinate matrix $\pmb E$ is given as
\begin{equation}
  {\pmb E}=
  \begin{pmatrix}
    \lambda_{s}\psi_{c}-\lambda_{c}\beta_{s}\psi_{s} & -\lambda_{s}\psi_{s}-\lambda_{c}\beta_{s}\psi_{c} & -\lambda_{c}\beta_{s} \\
    -\lambda_{c}\psi_{c}-\lambda_{s}\beta_{s}\psi_{s} & \lambda_{c}\psi_{s}-\lambda_{s}\beta_{s}\psi_{c} & -\lambda_{s}\beta_{c} \\
    \beta_{c}\psi_{s} & \beta_{c}\psi_{c} & -\beta_{s}
  \end{pmatrix},
\end{equation}
where the subscripts $_{.s}$ and $_{.c}$ are sine and cosine operations, respectively, the Galactic coordinates are latitude $\beta$ and longitude $\lambda$, and $\psi$ is the rotation around the direction of GW propagation; As the effect of varying the $\psi$ values is not the focus of this paper, to simplify computation, we have set $\psi$ to 0.

Many Galactic binaries have a frequency within the LISA frequency band and do not merge; however, some tight orbit systems show frequency variation. The computation for systems with frequency variation is more complex. For simplicity, ignoring systems that have frequency variation within the four-year LISA mission time, here we assume a monochromatic signal. We then have
\begin{equation}
\label{hpluscross2}
\left(\begin{array}{l}
h_{+}(t) \\
h_{\times}(t)
\end{array}\right)=\left(\begin{array}{l}
A_{+}(f) \cos \left(2 \pi f t+\phi_{0}\right) \\
A_{\times}(f) \sin \left(2 \pi f t+\phi_{0}\right)
\end{array}\right).
\end{equation}
The change in frequency or drift frequency $\dot{f}$ is small over the LISA mission time and ignored in our computational model; though studies, such as \cite{2022PhRvL.128d1101S}, suggest that it is possible to measure the drift frequency of Galactic ultra-compact binaries. However, for most systems $\dot{f}$ is mall unless $f$ is large. \citet{2019MNRAS.490.5888L} suggest that in a four-year mission that 750 sources may have detectable $\dot{f}$ and thus have measurable chirp masses. Importantly in our analysis it will not change the number of detected systems because for systems when $\dot{f}$ is high the sources are well above the sensitivity curve. This is consistent with other studies such as, \citet{2020ApJ...901....4B} and \citet{2022MNRAS.511.5936K}. To simulate the signal, for simplicity, we consider the phase $\phi_0$ is drawn from a uniform distribution between 0 and $2\pi$.

\begin{figure}
    \centering
    \includegraphics[width=60mm,scale=0.4]{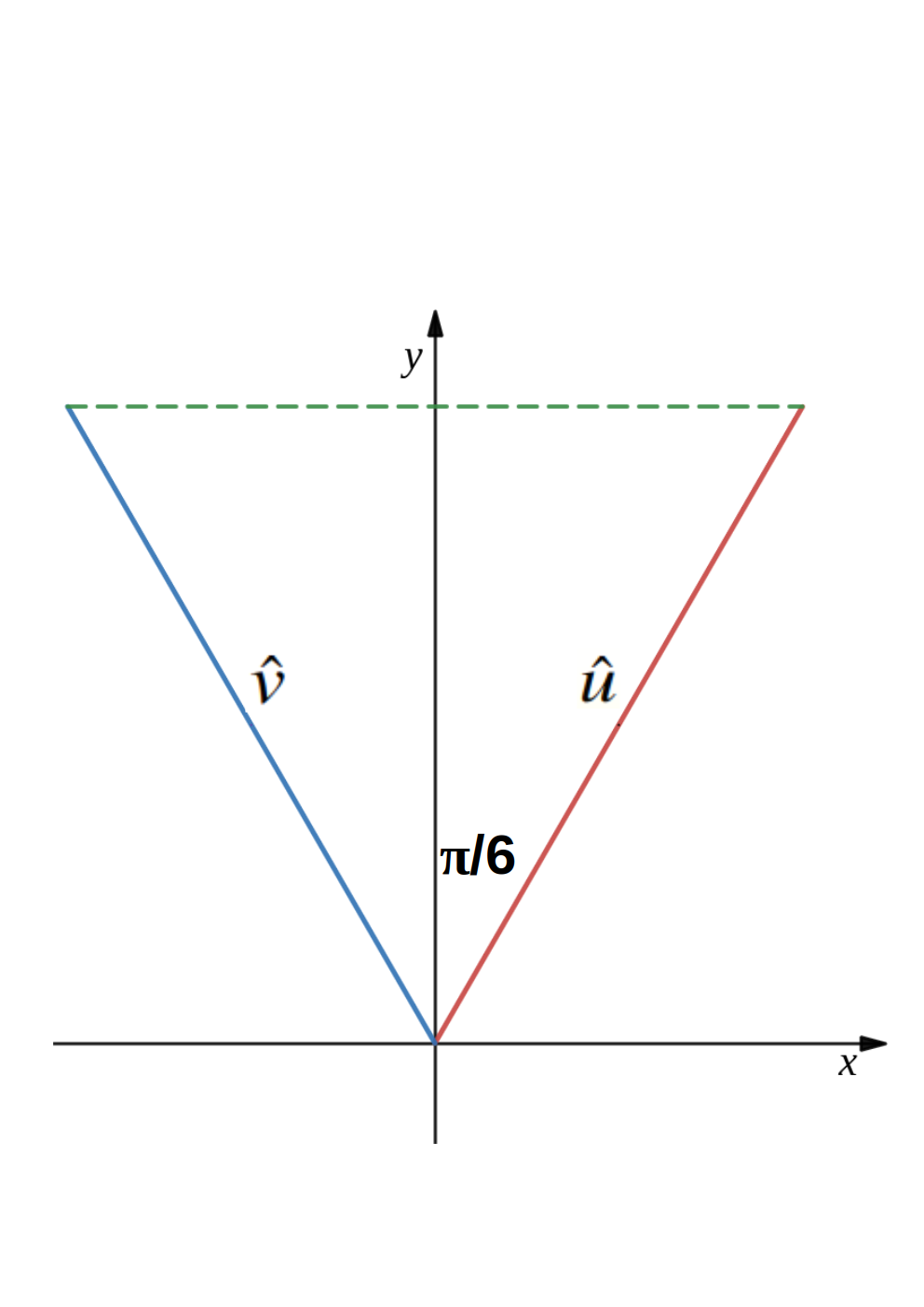}
    \caption{LISA coordinate system $(\hat{u}, \hat{v})$ used to evaluate the transfer function which is discussed in \protect \cite{2001CQGra..18.3473C}. LISA spacecrafts are located at the triangle's vertices, and all three arms are equal length.}
    \label{fig:LISA2}
\end{figure}

Depending on the location of the GW source in the sky, the signals detected by LISA are modulated. To calculate the response of the LISA detector arms, we use \citep{2007PhRvD..76h3006C}
\begin{equation}
\begin{aligned}
&\mathcal{H}_{+}(t)=A_{+}(f) \cos \left(2 \pi f t+\phi_{0}\right) {\pmb e}_{+}: \mathbf{D} \\
&\mathcal{H}_{\times}(t)=A_{\times}(f) \cos \left(2 \pi f t+\phi_{0}\right) {\pmb e}_{\times}: \mathbf{D},
\end{aligned}
\end{equation}
where the geometry dependent $\mathbf{D}$ is the one-arm detector tensor
\begin{equation}
\mathbf{D}=\frac{1}{2} (\hat{u} \otimes \hat{u}-\hat{v} \otimes \hat{v})
\end{equation}
and $\hat{u}$ and $\hat{v}$ are coordinate systems of the unit vectors along each arm of LISA spacecraft, defined as
\begin{equation}
  \hat{u}=\left(\begin{array}{c}
1 / 2 \\
0 \\
\sqrt{3} / 2
\end{array}\right)\quad
  \hat{v}=\left(\begin{array}{c}
-1 / 2 \\
\sqrt{3} / 2 \\
0
\end{array}\right).
\end{equation}
The sky-location-dependent polarised detector response functions outlined in \cite{2001PhRvD..65b2004C} are
\begin{equation}
\begin{aligned}
&F_{+}=-\frac{\sqrt{3}}{4}(1+\cos (\theta))^{2} \sin (2 \phi) \\
&F_{\times}=-\frac{\sqrt{3}}{2} \cos (\theta) \cos (2 \phi),
\end{aligned}
\end{equation}
where the source location $\phi$ and $\theta$ are related to the angles of the LISA transfer function as discussed in \citet{2001CQGra..18.3473C}. They are related via $\hat{u} \cdot \hat{z}=\sin(\phi+\pi /6)\sin \theta$ and $\hat{v} \cdot \hat{z}=\sin (\phi-\pi / 6) \sin \theta$, where $\hat{z}$ is the z direction in the Cartesian coordinate system. The LISA coordinate system is shown in Fig~\ref{fig:LISA2}. We assume that $\theta$ and $\phi$ change linearly and continuously with time in our computation.

Finally, we multiply the polarised strain and response function to generate a modulated signal from one source. To generate a Galactic binary signal from the whole population with N systems, we sum up all linear combinations of the two polarisations of the sources to get
\begin{equation}
\begin{aligned}
\label{eqn:modulated}
s(t)=& \sum_{i=1}^{N} \sum_{A=+, \times} h_{A, i}\left(f_{o r b, i}, M 1_{i}, M 2_{i}, X_{i}, Y_{i}, Z_{i}, t\right) \\
& \times F_{A}(\theta, \phi, t) \mathbf{D}(\theta, \phi, t) \end{aligned}
\end{equation}
or
\begin{equation}
\begin{aligned}
s(t)=& \sum_{i=1}^{N} \sum_{A=+, \times} \mathcal{H}_{A, i}\left(t\right) \times F_{A}(\theta, \phi, t).
\label{combinedmodulatedsignal}
\end{aligned}
\end{equation}
The full waveform will be used to calculate the modulated signal in later sections, while the predicted strain can be compared to LISA's sensitivity to determine its detectability.

\section{LISA Sensitivity Curve}
\label{noisecurve}
One way to show LISA's ability to detect GW sources is to compare the amplitude spectral density of the GW signals with the LISA full sensitivity curve $S(f)$ in the frequency domain, which is given in \cite{2019CQGra..36j5011R} as
\begin{equation}
S(f)=S_n(f) + S_{c}(f).
\end{equation}
A full sensitivity curve is the combined PSD of the LISA instrumental noise $S_n$ and the Galactic foreground confusion noise $S_c$. $S_n$ can be defined as
\begin{equation}
    S_n(f) = \frac{P_n(f)}{\mathcal{R}(f)},
\end{equation}
where $P_n(f)$ is the power spectral density (PSD) of the detector noise and $\mathcal{R}(f)$ is the sky-averaged detector response function in the frequency domain. $R(f)$ is defined as:
\begin{equation}
\mathcal{R}(f)=\frac{3}{10} \frac{1}{\left(1+0.6\left(f /f_*\right)^2\right)},
\end{equation}
where $f_*=c /(2 \pi L)$ is the transfer frequency \citep{2002PhRvD..66l2002P} and L is the LISA arm length of 2.5 million km. It is assumed that all three arms are equal in length. $P_n(f)$ is formed from the approximated single-link optical metrology noise $P_{\mathrm{oms}}$,
\begin{equation}
P_{\mathrm{oms}}=\left(1.5 \times 10^{-11} \mathrm{~m}\right)^2 \mathrm{~Hz}^{-1},
\end{equation}
and the single-test-mass acceleration noise $P_{\mathrm{acc}}$,
\begin{equation}
P_{\mathrm{acc}}=\left(3 \times 10^{-15} \mathrm{~m} \mathrm{~s}^{-2}\right)^2\left(1+\left(\frac{0.4 \mathrm{mHz}}{f}\right)^2\right) \mathrm{Hz}^{-1}.
\end{equation}
The above two types of instrument noises are then combined to form the total LISA instrument noise in a Michelson-style LISA data channel as
\begin{equation}
P_n(f)=\frac{P_{\mathrm{oms}}}{L^2}+2\left(1+\cos ^2\left(f / f_*\right)\right) \frac{P_{\mathrm{acc}}}{(2 \pi f)^4 L^2}.
\end{equation}

To formulate the full LISA sensitivity curve, we combine $S_n$ and the Galactic confusion noise $S_{c}$ from Galactic WDBs. The signal is modulated by the change in LISA's antenna response as the detector rotates over the LISA mission time. An averaged sensitivity curve approximation is made for our calculation; for an alternative approximation see \cite{2021PhRvD.104d3019K}. For plotting purposes we define the confusion noise as \cite{2019CQGra..36j5011R},
\begin{equation}
S_c(f)=Af^{-7 / 3} \mathrm{e}^{-f^\alpha+\beta f \sin (\kappa f)}\left[1+\tanh \left(\gamma\left(f_k-f\right)\right)\right] \mathrm{Hz}^{-1},
\end{equation}
where the amplitude A is 9 $\times$ $10^{-45}$ and the parameters of analytic fit for a four-year LISA mission time are $\alpha = 0.138$ , $\beta = -221$, $\kappa = 521$, $\gamma = 1680$ and $f_{k} = 0.00113$ \citep{2019CQGra..36j5011R}. It is important to appreciate that in our work, as we are modelling the WDB signal, the confusion noise from these binaries is model dependent, as we show later. However, we can show that individual binaries will be above this model limit. Each GW frequency (twice the orbital frequency) in our models is unique to a particular binary, and we assume it does not change over the LISA mission time (very ideal cases). A strain signal from a Galactic binary, including e.g. any of the verification binaries, is calculated as the one-sided, angle-averaged PSD $S_{h}(f)$
\begin{equation}
S_h(f)=\frac{A^2(f)}{2 T}, 
\end{equation}
where $A(f)$ is the amplitude of the wave and T is the LISA observation time frequency domain, in our case T is four years. $A(f)$ is related to distance between LISA and the signal source, the frequency and the $\mathcal{M}_c$ of the binary system. In later section, we defined the polarised strain which we will use to approximate signals in the time domain. For plotting in the frequency domain, we defined and computed the $A(f)$ using Eq.20--Eq.23 in \cite{2019CQGra..36j5011R}. 

The amplitude spectral density (ASD) is widely used by the ground-based detector community, for LISA we define the polarisation-averaged ASD as
\begin{equation}
ASD(f)= \sqrt{S_h(f)}. 
\end{equation}

Each individual ASD of a Galactic binary source is compared with the LISA sensitivity and known verification binaries in this section. To compute and plot $S_n$ and $S_{h}$ for the Galactic binaries and verification binaries, we use the Eq.1 - Eq.9 within the waveform model \textsc{PhenomA}  \citep{2007CQGra..24S.689A} and additional Python modules \citep{2019CQGra..36j5011R}. There are more generic and accurate models, such as the latest \textsc{PhenomB} model \citep{2014GReGr..46.1767H, 2015PhRvD..91b4043S}, which calculates spin-precession effects. \textsc{PhenomA} , however, is sufficient for calculating strains and estimating signal-to-noise ratios (SNRs) in our study.

\section{Calculation of SNR}
\label{snr}

The SNR detected by the LISA mission is affected by several factors. Some of the critical factors are as follows: Firstly, the SNR is directly proportional to the amplitude of the GW signal (for systems with known frequencies). Therefore, stronger signals produce higher SNR. Visually, the SNR of each  binary is given by the ratio of the height of the sensitivity curve and the height of the corresponding ASD in Figure~\ref{fig:bhbh_strain} and~\ref{fig:wdwd_phenoma}. The LISA mission is sensitive to GWs in the frequency range of 0.1 mHz to 1 mHz. The SNR is higher for signals that fall within this frequency range. Also, the longer the signal persists in the detector's bandwidth, the higher the SNR will be; hence a longer mission time increases the value of SNR. The SNR is also inversely proportional to the noise level of the detector. The lower the noise level, the higher the SNR will be. Finally, the distance and sky-location of the source also affect the SNR. The closer the source, the higher the SNR will be. Within our models, we can estimate the expected SNR for a given GW signal in the LISA mission.

A binary is detected if the SNR integrated over its observation time (in our case, we only consider the observation time of four years) is larger than an assumed threshold of 7. As the sky-location can affect the SNR for a source \citep{2022PhRvL.128d1101S}, without including the effect of sky-location, we calculate the SNR of a quasi-circular, non-spinning comparable-mass binaries with only a dominant quadrupole harmonic \citep{2019CQGra..36j5011R} of every binary as 
\begin{equation}
\overline{\rho^2}=\frac{16}{5} \int_{f_{\mathrm{in}}}^{f_{\mathrm{in}}+\Delta f} \frac{A^2(f)}{S_n(f)} \mathrm{d} f \approx \frac{16}{5} \frac{\Delta f A^2\left(f_{\mathrm{in}}\right)}{S_n\left(f_{\mathrm{in}}\right)},
\end{equation}
where $f_{in}$ is the initial GW frequency of the binary. Since we assume sources are monochromatic signals over the LISA mission time, the change in frequency $\Delta f$ can be ignored. We can simplify the SNR to form an approximation as

\begin{equation}
\label{fig:25}
\overline{\rho^2} \approx \frac{h_{\mathrm{GB}}^2\left(f_{\mathrm{in}}\right)}{S_n\left(f_{\mathrm{in}}\right)},
\end{equation}
where $h_{GB}$ is the strain spectral density in $\rm Hz^{-1/2}$, which can be expressed as
\begin{equation}
h_{\mathrm{GB}}=\frac{8 T^{1 / 2}\left(G \mathcal{M}_c / c^3\right)^{5 / 3} \pi^{2 / 3} f_{\text {in }}^{2 / 3}}{5^{1 / 2}\left(R / c\right)},
\end{equation}
where R is the distance between the GW source and the Sun.

For more complicated sources with known sky-location and inclination, \cite{2019CQGra..36j5011R} suggest a new expression in their Eq. (36), which calculates SNRs as for the sky-averaged case. Constant SNR scales with $\mathcal{M}_c$ to $\mathcal{M}_c^{5/3}$ in a given LISA’s volume sensitivity, and an SNR threshold of 7 which is computed assuming a four-year LISA mission duration with 100 per cent duty cycle and for (quasi-)stationary equal-mass circular binaries \citep{2023LRR....26....2A}. LISA will see Sources with SNRs > 7. Extreme mass ratio (EMR) binaries, which may be highly eccentric, require a more involved formalisation. Even for binaries with spins, the SNR we compute should be accurate to within a factor of two or so for spinning systems according to \cite{2019CQGra..36j5011R}.

\section{Extra Properties of the BPASS binary populations}

\begin{figure*}
    \centering
    \includegraphics[width=2\columnwidth]{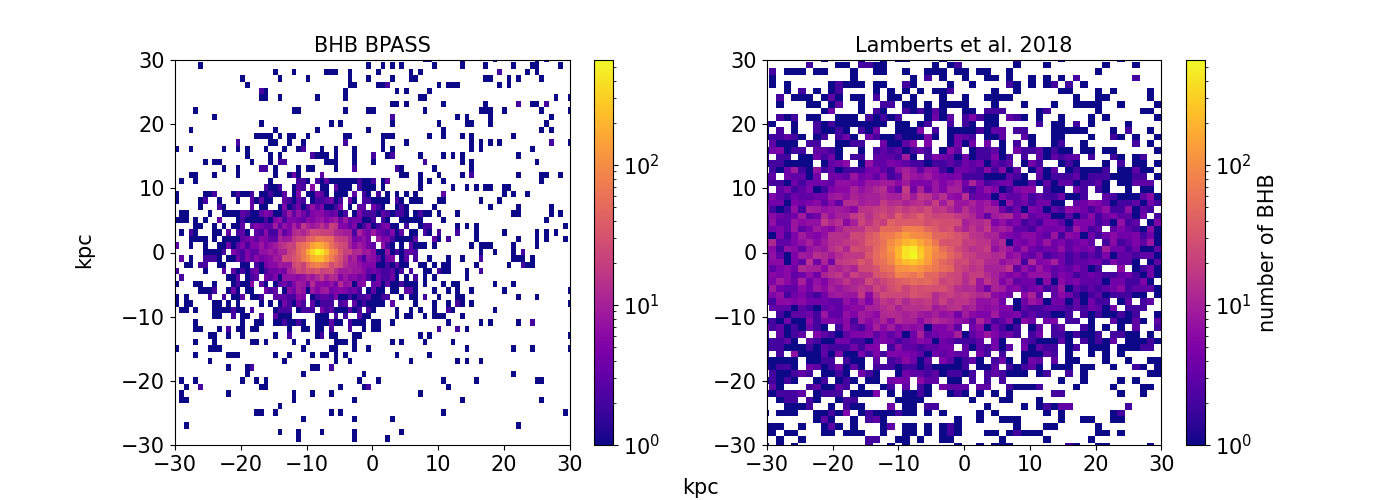}
    \caption{ The small-scale 2D number density map of BHB populations in the simulated galaxy from the two model predictions viewed face on. BHBs are present in Galactic bulge and disks in both model predictions. The population density is the highest at the centre of both galaxies.}
    \label{fig:number_bhbh_short}
\end{figure*}
\begin{figure*}
    \centering
    \includegraphics[width=2\columnwidth]{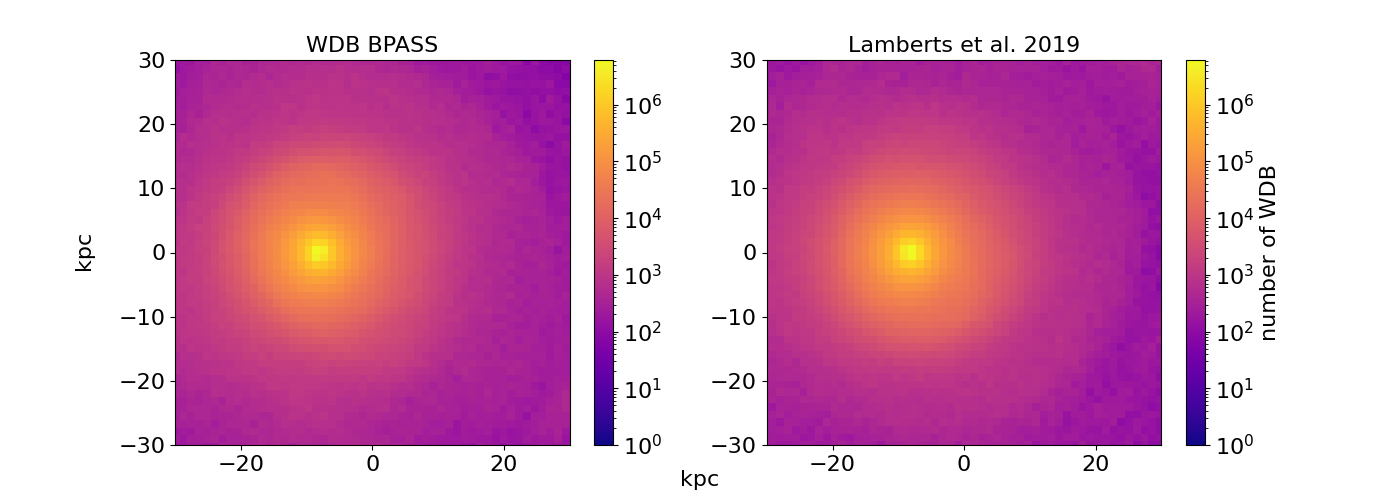}
    \caption{ The small-scale 2D number density map of WDB populations in the simulated galaxy from the two model predictions viewed face on. WDBs are present in the Galactic bulge and disks in both model predictions. The population density is the highest at the centre of both galaxies.}
    \label{fig:number_wdwd_short}
\end{figure*}
Figures~\ref{fig:number_bhbh_short} and~\ref{fig:number_wdwd_short} show small-scale 2D density maps of the comparison between BPASS and BSE for Galactic BHB and WDB populations. The BPASS BHB population is less spread out in the MW than the BSE BHB population, and on the other hand, the BPASS WDB populations are similarly distributed in the MW galaxy.

\begin{figure*}%
    \centering
    \includegraphics[width=\columnwidth]{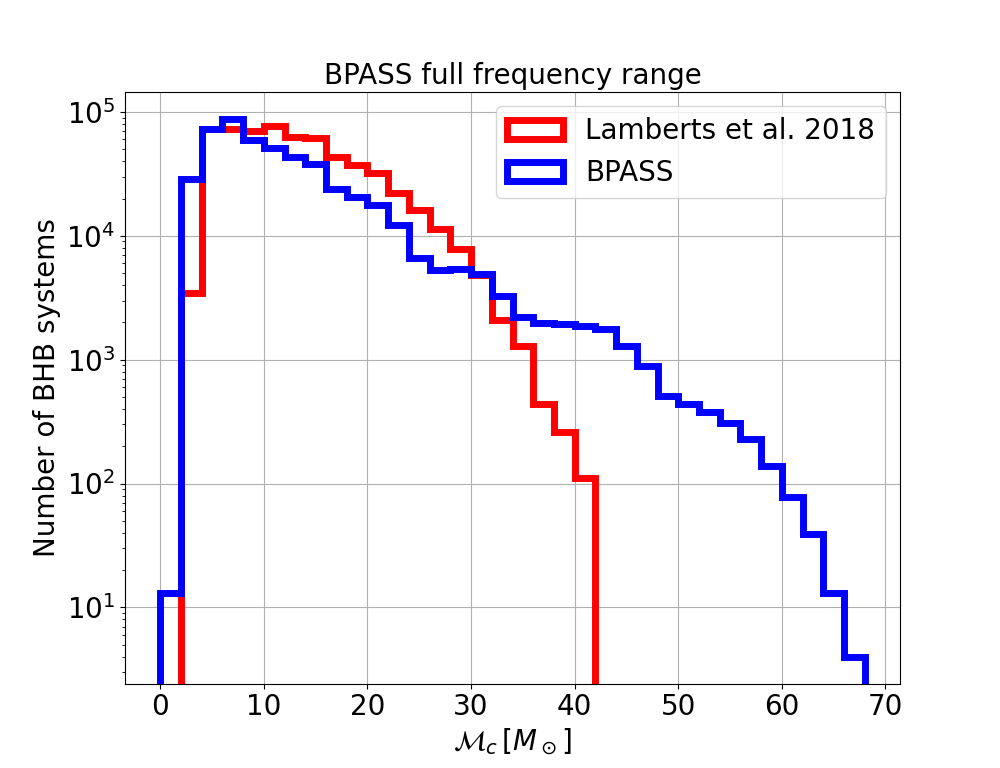}
    \includegraphics[width=\columnwidth]{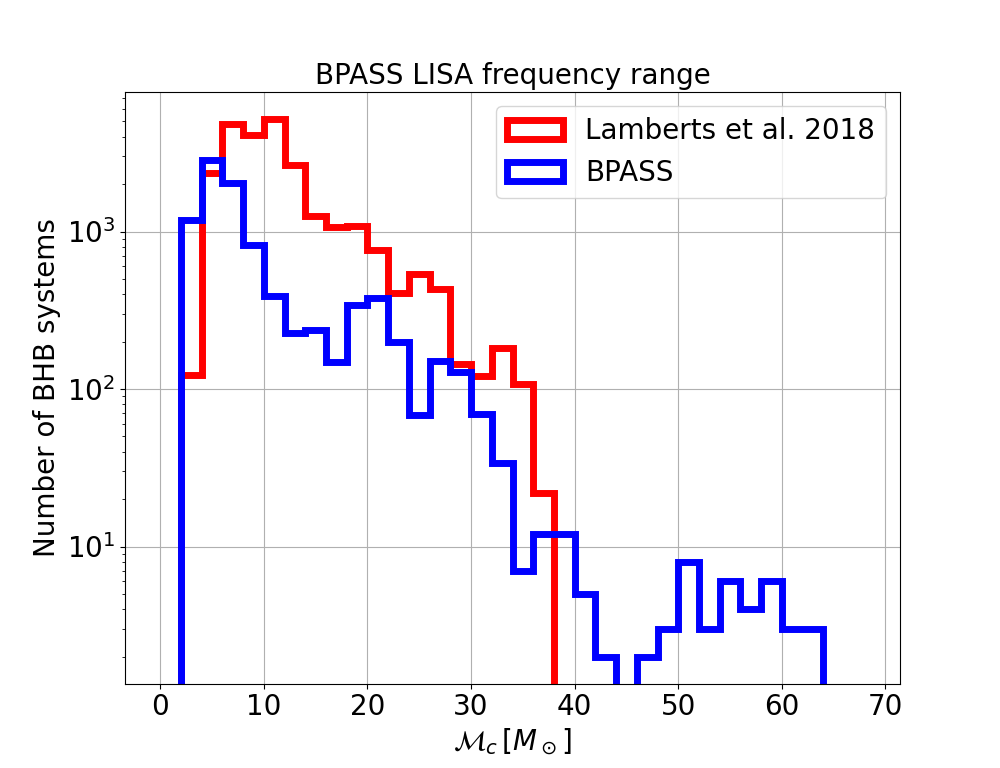} 
    \caption{The distribution of BHB by $\mathcal{M}_c M_\odot$. The left panel includes the distribution of BHBs over the full frequency range of $10^{-10}$ Hz--$0.1$ Hz, the right panel contain the frequency range ($10^{-5}$ Hz--$0.1$ Hz).}
    \label{fig:bhbh_Mc}%
\end{figure*}
\begin{figure*}%
    \centering
    \includegraphics[width=\columnwidth]{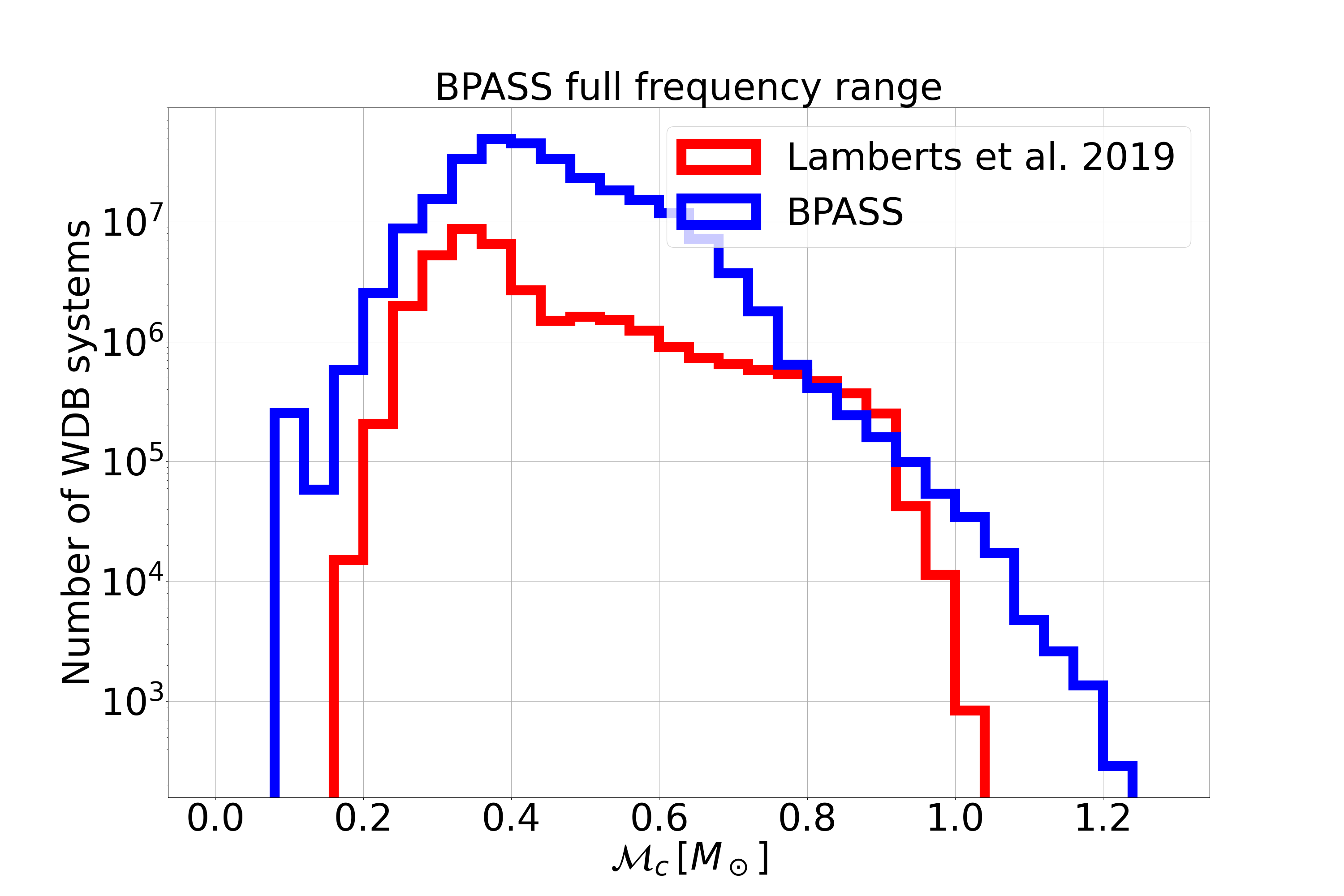}
    \includegraphics[width=\columnwidth]{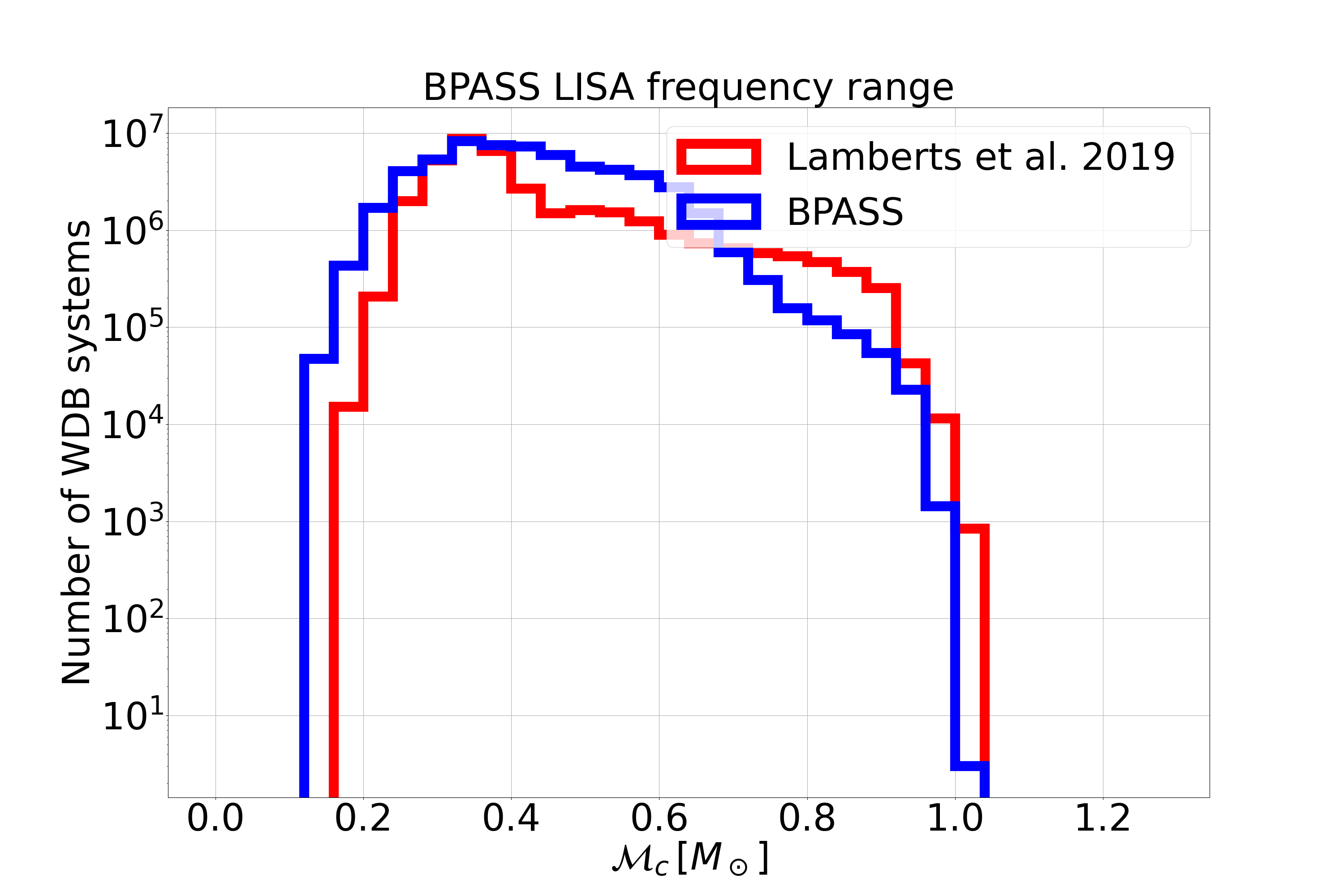}
    \caption{The number distribution of WDB by $\mathcal{M}_c M_\odot$. In both panels the BSE results only include the frequency range $10^{-5}$ Hz--$0.1$ Hz. However on the left panel the BPASS results incldue the full frequency range of $10^{-10}$ Hz--$0.1$ Hz, the right panel contain the frequency range $10^{-5}$ Hz--$0.1$ Hz.}%
    \label{fig:wdwd_Mc}
\end{figure*}

Most of the BPASS BHBs with $\mathcal{M}_c$ greater than 40$M_\odot$ are outside the frequency range of LISA, shown in Figure~\ref{fig:bhbh_Mc}, BPASS predicts a total number of 495,586 BHBs in the MW galaxy and only 9,298 (1.9 per cent) of the total BHB population are within the LISA frequency band, compared with \cite{2018MNRAS.480.2704L} who found 4.2 per cent within the band. On the other hand, the BPASS predicts a total number of 272,789,350 WDBs and 58,522,007 of these are within the LISA frequency band, nearly 21.5 per cent. Shown in Figure~\ref{fig:wdwd_Mc} range, the population of these 'non-LISA' WDBs distributes evenly across the $\mathcal{M}_c$, and they contribute to the confusion noise.

\begin{figure*}%
    \centering
    \includegraphics[width=\columnwidth]{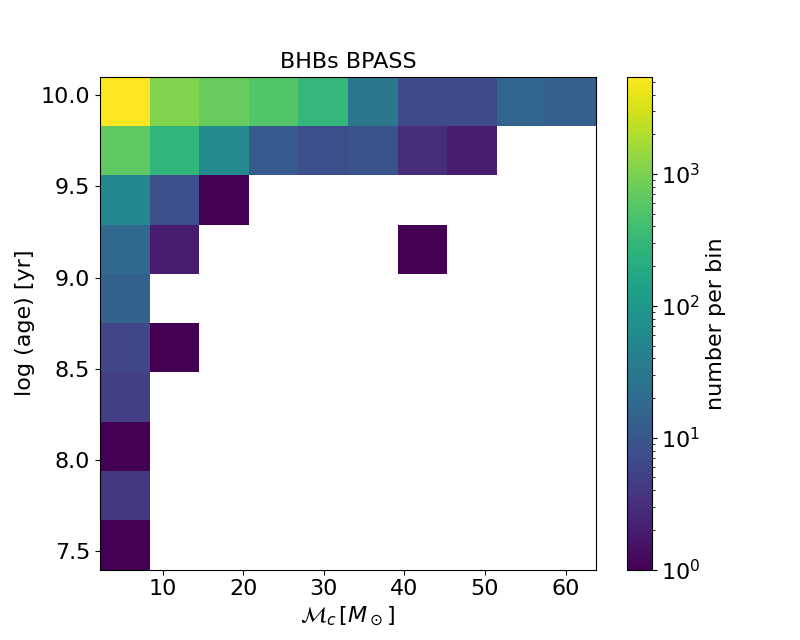} %
    \includegraphics[width=\columnwidth]{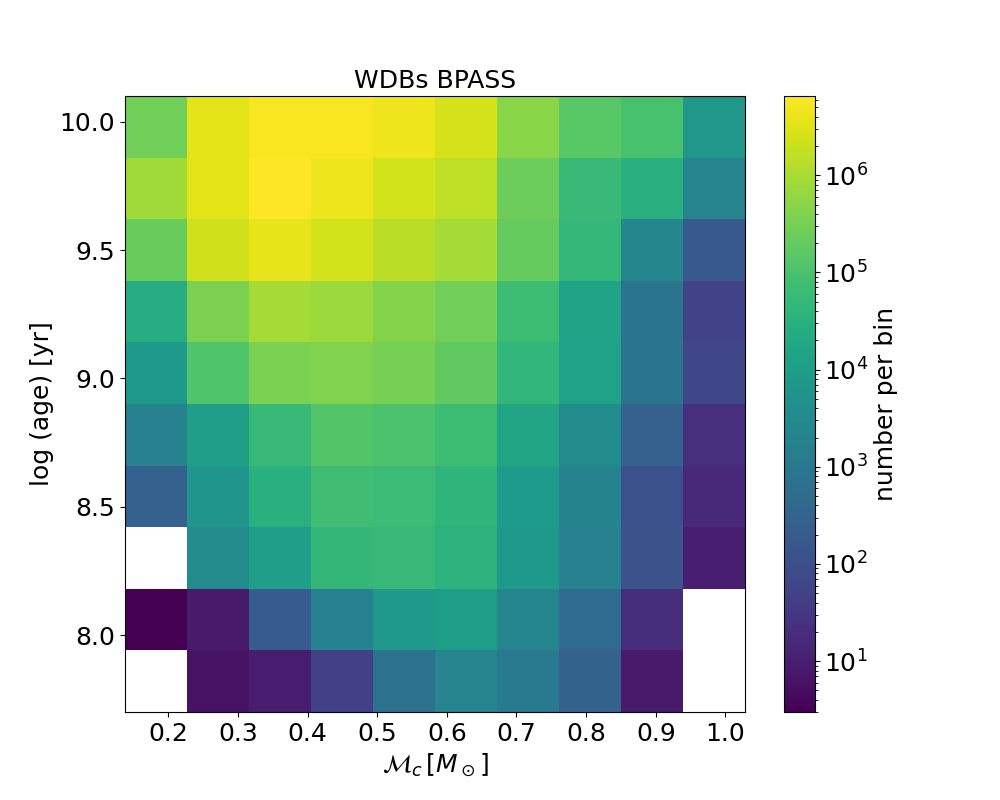} %
    \caption{The BPASS BHBs and WDBs distribution in $\mathcal{M}_c$ versus age distribution.}%
    \label{fig:age}%
\end{figure*}
We take another look at the BPASS BHB and WDB populations in the LISA frequency band in Figure~\ref{fig:age}; where most of the BHBs have $\mathcal{M}_c$ less than 10$M_\odot$ and are about 10Gyr old. Similarly, older WDBs with $\mathcal{M}_c$ below 0.6$M_\odot$ are more numerous in the WDB population.

\begin{figure*}%
    \centering    
    \includegraphics[width=\columnwidth]{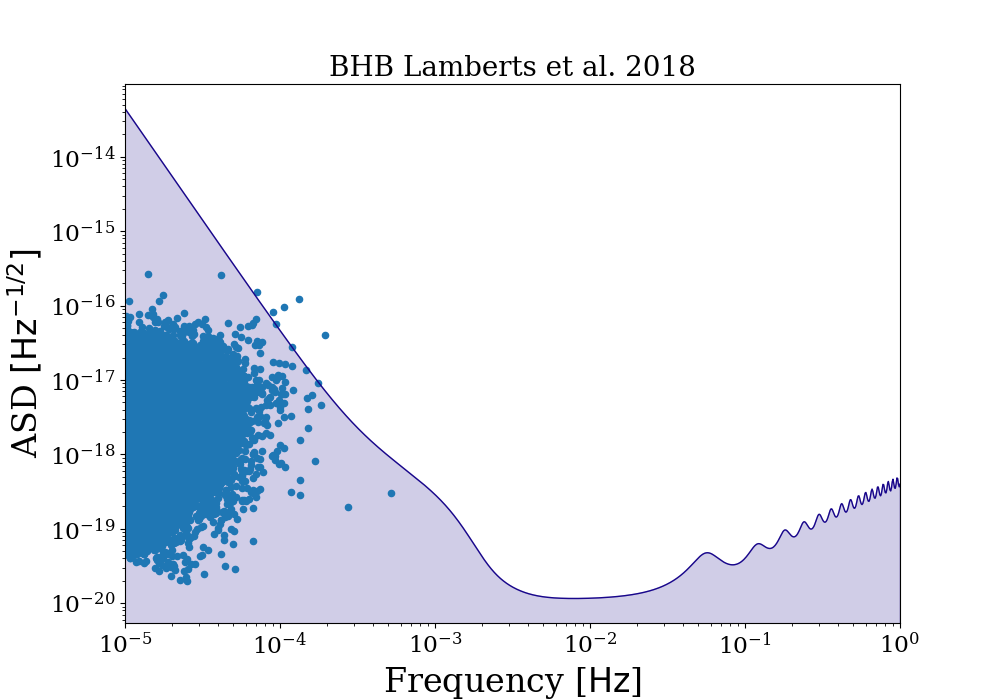} %
    \includegraphics[width=\columnwidth]{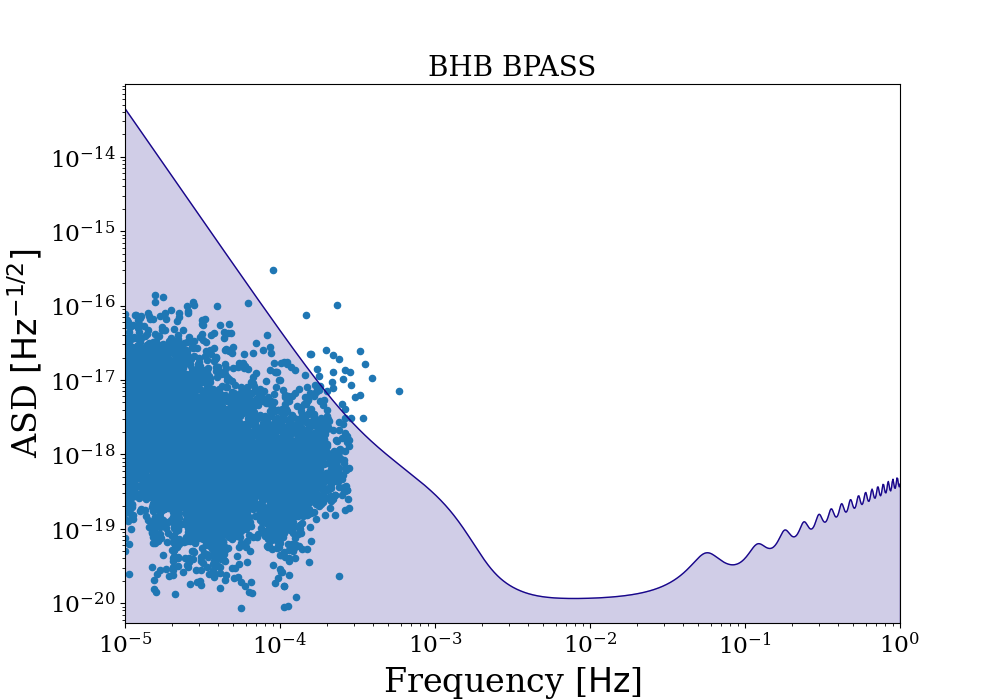} %
    \caption{Same as in Figure \ref{fig:bhbh_strain} but now using \textsc{LEGWORK}. The extra sources above the LISA sensitivity curve may be caused by different inclination assumptions. The inclination is not taken into account in \textsc{LEGWORK}. However, in \textsc{PhenomA}, we have generated random inclinations to calculate the waveform.}%
    \label{fig:legwork}%
\end{figure*}
\begin{figure*}
    \centering
    \includegraphics[width=\columnwidth]{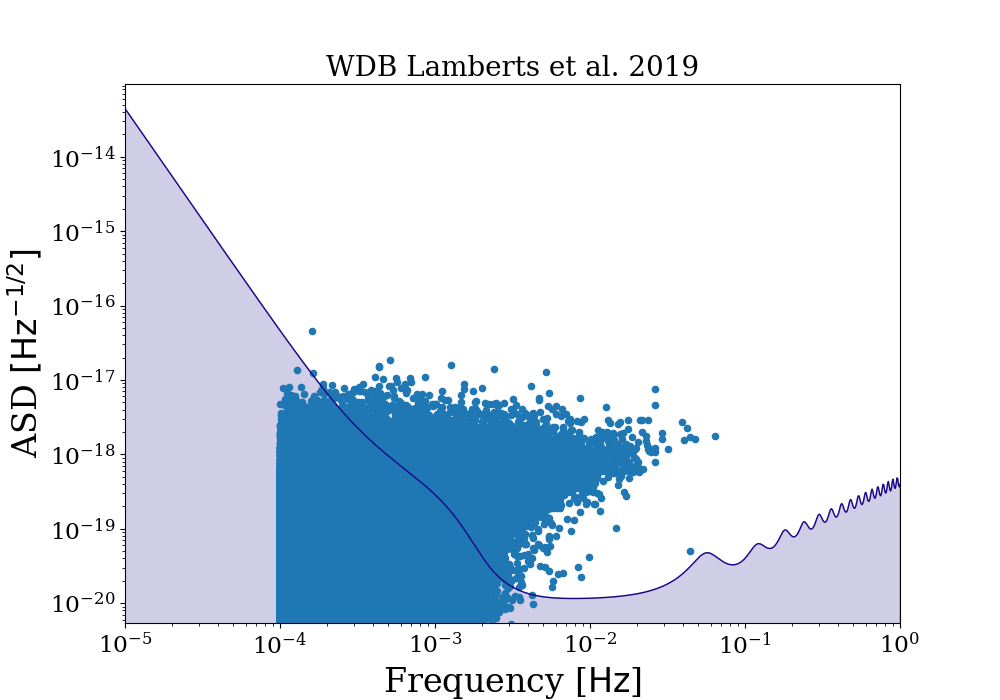} %
    \includegraphics[width=\columnwidth]{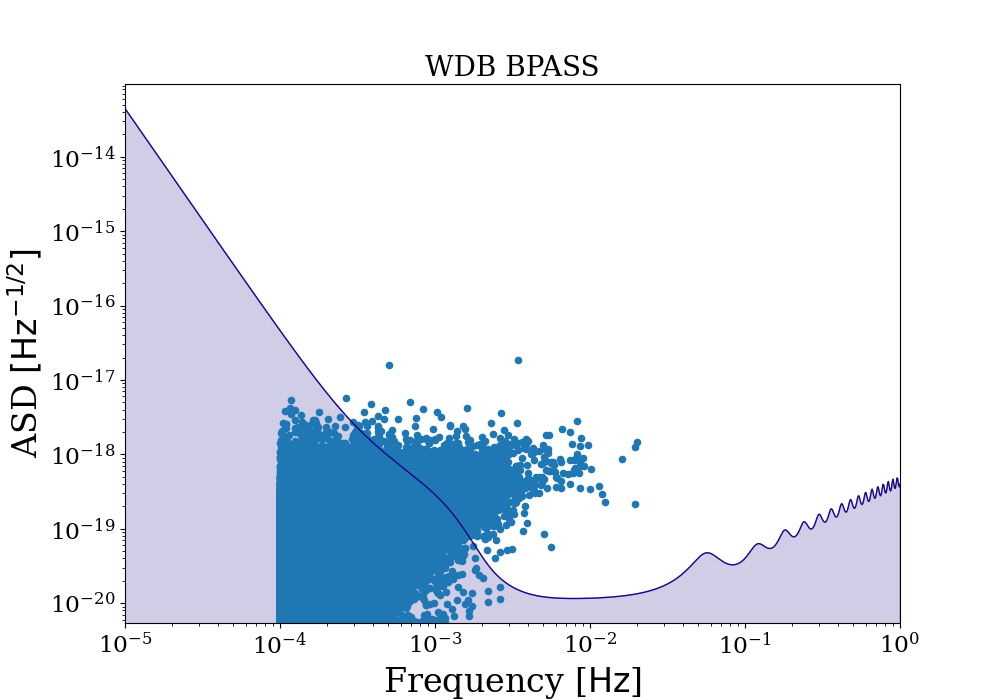}%
    \caption{Same as in Figure~\ref{fig:wdwd_phenoma} but now using \textsc{LEGWORK}. The extra sources above the LISA sensitivity curve may be caused by different inclination assumptions. The inclination is not taken into account in \textsc{LEGWORK}. However, in \textsc{PhenomA}, we have generated random inclinations to calculate the waveform.}
    \label{fig:wdwd_strain}
\end{figure*}

We provide an alternative frequency vs ASD plot using LEGWORK in Figure~\ref{fig:legwork}. Compared with Figure~\ref{fig:bhbh_strain} and~\ref{fig:wdwd_strain}, we notice a shift in the scale of the ASD. This is because of the difference in the ASD formulation. The alternative ASD defined in LEGWORK is
\begin{equation}
\rm    ASD=h_2 \sqrt{T}
\end{equation}
where $h_2$ is the strain amplitude of the source for the second orbital frequency harmonic \citep{2022ApJS..260...52W,2023ApJ...945..162T}. Both BSE and BPASS BHB populations are similar with a small fraction of the population above the LISA sensitivity curve at lower frequencies. Unlike the results from the \textsc{PhenomA} , here the high-frequency WDBs in both BSE and BPASS models have ASDs of about $\rm 10^{-17} Hz^{-1/2}$.

We present an alternative WDB population f vs strain plot using LEGWORK in Figure \ref{fig:wdwd_strain}. Similar to the \textsc{PheomA} results, the BSE WDB population have similar strains and more systems at a higher frequency than BPASS.

\begin{figure*}
    \centering
    \includegraphics[width=2\columnwidth]{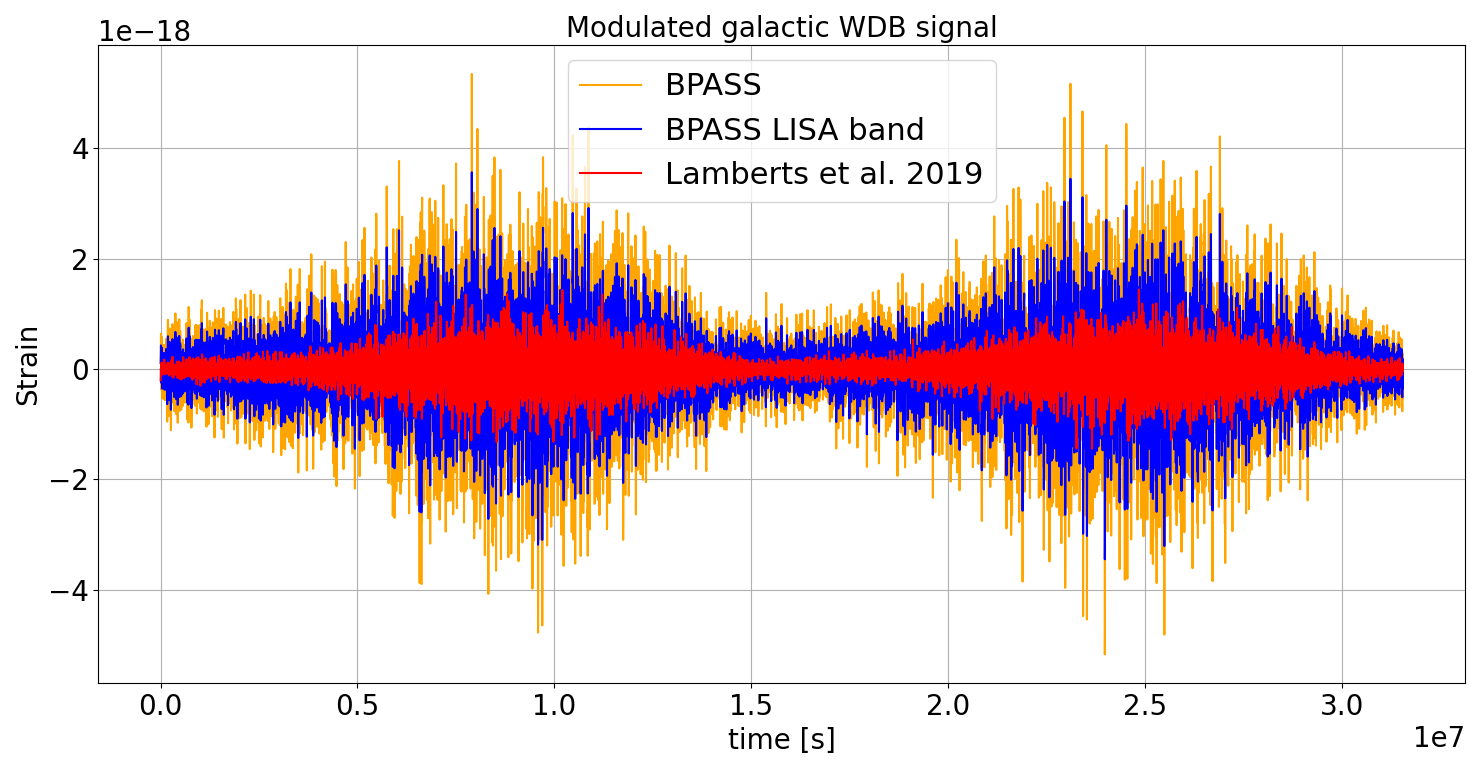}
    \caption{ Same as in Figure \ref{fig:modulated_wdwd} but now including a yellow line that is the total signal from BPASS WDBs with frequency ranging $10^{-10} - 10^{1}$ Hz}
    \label{fig:modulated_wdwd3}
\end{figure*}

Modulated signal observed by LISA is the superposition of many binary sources. BPASS WDBs in the LISA frequency contribute to almost all the signal strength that are seen by LISA,  shown in Figure~\ref{fig:modulated_wdwd3}, this could be because the 'non-LISA' BPASS WDBs are less massive and/or really far from us.

\bsp
\label{lastpage}
\end{document}